\documentclass[11pt,a4paper]{article}
\pdfoutput=1
\usepackage{jheppub}
\usepackage{amsmath,amscd}
\usepackage{amsfonts}
\usepackage{amssymb}
\usepackage{graphicx}
\usepackage{color}
\usepackage[normalem]{ulem}
\usepackage{hyperref}
\usepackage{linearb}

\newcommand{\RR}{\mathbb{R}} 
\newcommand{\ZZ}{\mathbb{Z}} 
\newcommand{\NN}{\mathbb{N}} 

\newcommand{\G}{\mathcal{G}}

\def\tr         {{\rm  tr}}
\def\cala         {{\cal A}}

\def\calc         {{\cal C}}

\def\calf         {{\cal F}}

\def\calh         {{\cal H}}

\def\caln         {{\cal N}}

\def\calu         {{\cal U}}

\def\calw         {{\cal W}}
\def\calz         {{\cal Z}}

\def\be{\begin{equation}}
\def\ee{\end{equation}}
\def\bea{\begin{eqnarray}}
\def\eea{\end{eqnarray}}

\def\a{\alpha}
\def\b{\beta}
\def\h{\eta}

\def\G{\Gamma}
\def\d{\delta}

\def\D{\Delta}
\def\l{\lambda}
\def\L{\Lambda}

\def\f{\phi}

\def\o{\omega}
\def\O{\Omega}
\def\p{\pi}
\def\r{\rho}

\def\s{\sigma}

\def\t{\tau}

\def\sF{{{ F}\!\!\!\!\hskip.8pt\hbox{\raise1pt\hbox{/}}\,}}
\def\som{{{ \omega}\!\!\!\!\hskip.8pt\hbox{\raise1pt\hbox{/}}\,}}
\def\sJ{{{\rm J}\!\!\!\!\hskip.8pt\hbox{\raise1pt\hbox{/}}\,}}

\def\NO{{\tiny\textlinb{\BNcc}}}
\def\F{\Phi}
\def\pa{\partial}

\def\to{\rightarrow}
\def\nonu{\nonumber \\{}}
\def\half{{1 \over 2}}

\def\red{\textcolor{red}}
\def\blue{\textcolor{blue}}

\title{On tensionless string field theory in AdS$_3$}

\author{Joris Raeymaekers}

\affiliation{CEICO, Institute of Physics of the Czech Academy of Sciences, \\ Na Slovance 2, 182 21 Prague 8, Czech Republic.}
\emailAdd{joris@fzu.cz}

\abstract{We report on progress in formulating a field theory of tensionless strings in $AdS_3$, starting from the dual large-$N$ symmetric orbifold CFT. We propose a set of field equations which are gauge invariant under the higher spin algebra of the theory, the  `Higher Spin Square'. The massless higher spin sector is captured by a Chern-Simons gauge field, while the matter sector is described by unfolded equations similar  to those appearing in Vasiliev theory. Our equations incorporate the full perturbative spectrum of the theory, including states coming from the  twisted sectors, and capture some of the interactions  fixed by gauge invariance. We also discuss the spectrum of the bulk theory and explain how linearization around $AdS_3$ gives rise to the expected  set of decoupled wave equations. Our results can be  generalized to describe bulk duals of other large-$N$ symmetric
	orbifolds.}
\arxivnumber{}
\keywords{}
\begin{document}
\maketitle

 \section{Introduction}
 To deepen our understanding of string theory, it is of obvious importance to clarify its underlying symmetry principle.
    When expanded  around  the Minkowski background, string theory takes the form of  a massive higher spin theory   containing towers of fields lying on Regge trajectories of increasing mass and spin. Its behaviour at large energies   suggests a huge underlying
higher spin gauge symmetry, which is spontaneously broken in the
Minkowski vacuum \cite{Gross:1988ue}. 
This symmetry, or portions thereof, is  expected to  become  manifest
 in tensionless limits of the theory. In such limits, the structure of string field  theory is expected to become much more transparent: all the fields should organize themselves in multiplets of the symmetry, and  gauge-invariance strongly constrains  the allowed interactions. 
 
 One of the lessons from the work of Vasiliev \cite{Vasiliev:1990en} is that higher spin gauge theories are naturally formulated on (anti-)de Sitter backgrounds; this suggests to make sense of  the tensionless limit  in an  $AdS$ vacuum of string theory \cite{Sundborg:2000wp}. This idea has recently been given a beautiful and  explicit realization in the work of Eberhardt, Gaberdiel and Gopakumar \cite{Eberhardt:2018ouy}. They considered the worldsheet\footnote{See \cite{Lindstrom:2003mg},\cite{Schild:1976vq,Isberg:1992ia,Isberg:1993av,Gustafsson:1994kr,Bonelli:2003kh,Bonelli:2003zu,Bagchi:2015nca} for a partial list of earlier works on the tensionless limit on the worldsheet.} theory of strings  propagating on the $AdS_3 \times S^3\times T^4$ background with Neveu-Schwarz flux; this background arises as the near-horizon limit of $k\ NS5$-branes and $N$ fundamental strings. The value of $k$ sets the ratio between the $AdS$ radius $l_{AdS}$  and the string length $l_s$, 
\be k = \left({l_{AdS} \over l_s}\right)^2, \label{kval}
\ee 
The smallest value of $k$  for which this background   makes sense is $k=1$, and this value corresponds to a  tensionless limit of the theory\footnote{A dual description for the  theory at $k>1$ was recently proposed in \cite{Eberhardt:2019qcl}.} \cite{Lindstrom:2003mg}. The string  coupling is proportional to $1/\sqrt{N}$, so $N$ should be taken to be large. The analysis of the worldsheet theory  at $k=1$ in \cite{Eberhardt:2018ouy} gave strong evidence, from the spectrum as well as the interactions, that this tensionless $AdS_3$ string is dual to the the  $N$-th symmetric orbifold 
 of the $\caln = (4,4)$ SCFT on  $T^4$. The tensionless character of the $k=1$ limit manifests itself in the fact that the dual theory is quasi-free and possesses a huge chiral algebra.

 The dual symmetric orbifold CFT   was of course already  proposed a long time ago  from considerations in the    S-dual D1-D5 frame \cite{Strominger:1996sh}. However in this frame the derivation  was  less direct;  it was  for example unclear where the symmetric orbifold is located  in the  20-dimensional moduli space of the theory. This  limited the checks
 of the proposal to the study of protected quantities \cite{deBoer:1998kjm,deBoer:1998us,Maldacena:1999bp} (see \cite{David:2002wn} for a review and further references).

 Our goal in the present work is to take a first step in understanding how the interactions in the tensionless string field theory on $AdS_3 \times S^3\times T^4$ are constrained by higher spin gauge invariance.
  While the recent understanding of the worldsheet may eventually make it possible to  use a conventional formulation of string field theory for this purpose (as was initiated in \cite{Sagnotti:2003qa} for open strings in flat space), we will here take a different route and use the boundary CFT  as our starting point for reconstructing the bulk theory. In the series of papers \cite{Gaberdiel:2014cha,Gaberdiel:2015wpo,Gaberdiel:2015mra}, Gaberdiel and Gopakumar derived
 properties of the bulk theory from the symmetric orbifold dual. The higher spin gauge algebra has the structure of 
a `Higher Spin Square' ($HSS$), containing both `horizontally' and `vertically' embedded conventional higher spin subalgebras, while the full algebra is in some sense exponentially larger. From the relation (\ref{kval}) we see that  it is not possible to take the string length $l_s$ all the way to infinity in the $AdS_3$ background, and therefore the  theory also contains massive matter fields. These arise from the untwisted as well as the twisted sectors in the CFT and fall in multiplets of the $HSS$ algebra.

We will propose a set of equations which describe the tensionless string field theory to linear order in the matter fields.
These equations take a form familiar from linearized Vasiliev theory \cite{Vasiliev:1992gr}:
\bea 
F = d A + A \wedge A &=&0, \qquad   \bar F = d \bar A + \bar A \wedge \bar A =0\nonu
\left( d + A^{(n)} + \bar A^{(n)}  \right)| C^{(n)}\rangle &=&0, \qquad  n= 1,2, \ldots\label{eqsintro}
\eea
Here, $A$ and $\bar A$ are one-form  fields taking values in $HSS$; they describe the massless higher spin fields. The vanishing of the field strengths in the first line  corresponds to the fact that these fields don't contain any local degrees of freedom, though as usual in $AdS_3$ they do describe boundary excitations. The second line is a set of  equations describing the matter sector in an `unfolded' formulation (see \cite{Bekaert:2005vh} for a review). Each $| C^{(n)}(x)\rangle $ is a  zero-form  taking values in  the  Hilbert space of a particular twisted sector of the orbifold CFT. When linearized around $AdS_3$ the equations describe, in a manner which we will make precise, an infinite number of  $AdS_3$ matter fields of arbitrary high spin. The superscript $(n)$ in  $A^{(n)}, \bar A^{(n)}$ denotes that the gauge
fields must be evaluated in the appropriate representation acting on the twisted sector Hilbert space; one of the main results of the paper is to make this precise.
The equation  for $n=1$, which captures matter coming from the untwisted sector, was already proposed in \cite{Raeymaekers:2016mmm}. 
The equations (\ref{eqsintro})  contain interactions coming from vertices
of the type $AAA$ and $CCA$   and are expected to capture the corresponding holographic three-point
functions  (see  \cite{Ammon:2011ua}).

To make this work self-contained, we will review some of the results of \cite{Gaberdiel:2014cha,Gaberdiel:2015mra,Gaberdiel:2015wpo}
needed for our analysis. The paper is organized as follows. Since most of what we have to say applies to general symmetric orbifolds at large $N$,  we start 
in section \ref{secsymmorb} with a  review of the spectrum of single-particle excitations in general symmetric orbifolds and the underlying Hilbert spaces.
In section \ref{secbulk} we give an equally general discussion of the proposed bulk dual higher spin theory, showing  how 
the requirements of capturing the CFT spectrum and of higher spin gauge gauge invariance lead to equations of the form (\ref{eqsintro}).
We also summarize the results of \cite{Kessel:2018zqm} which allow us  to eliminate the auxiliary components  in the unfolded fields $|C^{(n)}\rangle $ and obtain  physical
fields satisfying the appropriate wave equations in $AdS_3$.  In section \ref{secscalar} we apply this general framework to a simple example, the symmetric orbifold of a free boson. We show
how the matter fields $| C^{(n)}\rangle$ can be described explicitly in terms of oscillators, and give an algorithm to evaluate the gauge fields 
$A^{(n)}, \bar A^{(n)}$ in the appropriate representation. In section \ref{sechss} we turn to our main example of interest, that of the $T^4$ symmetric orbifold. We address a subtlety in  treating the fermionic sector and focus on making the ingredients entering in eqs. (\ref{eqsintro}) precise. We also include an analysis of the spectrum of supermultiplets described by our equations. We finish by listing some open problems and future directions.

 \section{Symmetric orbifolds at large $N$}\label{secsymmorb}
 Large-$N$ symmetric orbifold CFTs provide us with a  class of holographic conformal field theories  with a `stringy' bulk dual, in the sense that the spectrum has Hagedorn-like behaviour \cite{Keller:2011xi}. The example of the $T^4$  symmetric orbifold actually describes  tensionless strings propagating on $AdS_3$, while other examples of symmetric orbifolds with a dual string theory interpretation appear in \cite{Giribet:2018ada,Gaberdiel:2018rqv,Eberhardt:2019qcl}.

 Large-$N$ factorization in holographic CFTs implies that the spectrum separates  into `single-particle' excitations, which are dual to normalizeable modes of elementary fields in the bulk dual, and `multi-particle' excitations. If the CFT is a large-$N$ gauge theory, these are the states created by single-trace and multi-trace operators respectively. For the symmetric orbifold
 CFTs at large $N$ it is not immediately obvious what is the equivalent of single-trace excitations.
 In this section we derive the single-particle spectrum of large-$N$ symmetric orbifolds by taking a limit of the partition function, following the analysis of \cite{Gaberdiel:2015mra,Gaberdiel:2015wpo}.
 In doing so we will also  keep track of the actual states counted by the partition function, which  comprise the single-particle Hilbert space; this space will play a crucial role in our subsequent  construction of the  dual field theory.

 For technical simplicity, we will focus here on symmetric orbifolds of bosonic theories, though the analysis generalizes to fermionic theories modulo a technical subtlety involving $NS$ sector fermions. Discussion of this issue will be postponed until  section \ref{sechss}. 
 \subsection{Symmetric orbifolds: generalities}\label{secdmvv}
 Here we review some general aspects  of symmetric orbifold CFTs, see \cite{Dijkgraaf:1996xw,Dijkgraaf:1998zd,Maldacena:1999bp} for more details.
 We start from a 2D bosonic `seed' conformal field theory $\calc$ 
with central charge $c$  whose Hilbert space we denote by $\calh$.
 The partition function 
of the seed theory is
 \bea 
 Z (\calh) &\equiv & \tr_{\calh } q^{L_0} \bar  q^{\bar L_0}  \nonu
 &=& \sum_{\D ,\bar \D} c(\D ,\bar \D ) q^\D \bar q^{\bar \D}.\label{ZC}
  \eea
  where we introduced the degeneracies $c (\D, \bar \D )$.
 We will assume the theory to be well-defined on a cylindrical worldsheet  of circumference $2\p$, which requires the quantization condition
  \be
  \D - \bar \D \in \ZZ, \qquad \forall\ \D, \bar \D \label{hminhbar}
  	\ee
  	on the spectrum.
  	We will take the convention that the conformal  vacuum in the theory on the cylinder  has weights
  	\be \D_0 = \bar \D_0 =  -{c \over 24}.
  	\ee
  	 Note that for simplicity we have assumed  the left- and right central charges to be equal, though this is not essential. It is sometimes convenient to consider the partition function with the overall factor $(q \bar q)^{-{c\over 24}}$
  	 removed; we will denote this partition function and its expansion coefficients with a tilde, i.e. 
  	 \be \tilde Z (\calh) \equiv  (q \bar q)^{{c\over 24}} Z = \sum_{h, \bar h} \tilde c(h, \bar h) q^h \bar q^{\bar h}.\label{ZCtilde} \ee
  In particular,  \be 
  h = \D - {c \over 24}, \qquad \bar  h = \bar \D - {c \over 24}, \qquad  \tilde c(h, \bar h)= c\left( -{c\over 24}+ h ,  -{c\over 24}+ \bar h\right).\ee
  	Unitarity of the theory implies that all $h, \bar h$ are positive.

 Now we want to consider a new CFT which is the 
 $N$-th symmetric product orbifold of $\calc$:
  \be
 Sym^N (\calc) \equiv {\calc^N\over S_N}.
  \ee
  Here the symmetric group $S_N$ acts by permuting  $N$ noninteracting copies of $\calc$. 
  The resulting CFT can be completely constructed from the seed theory $\calc$ \cite{Dijkgraaf:1996xw,Dijkgraaf:1998zd,Maldacena:1999bp}. The general theory of orbifolds \cite{Dixon:1985jw,Dixon:1986jc} tells us  that the Hilbert 
  space of this orbifold CFT is of the form
  \be
  \calh \left(  Sym^N (\calc) \right) = \bigoplus_{[g]} \calh_g^{C_g}
  \ee
  Here, the sum runs over twisted sectors which are labelled by conjugacy classes of $S_N$ and  $C_g$ is the centralizer subgroup of a representative $g$ within this conjugacy class. The notation  $\calh^C$ means that we project on the subspace of $\calh$ invariant under the action of $C$.

  The  conjugacy classes of $S_N$ are in one-to-one correspondence with the partitions of $N$ and can be labelled as
  \be
 [g] = (1)^{N_1} (2)^{N_2} \ldots (f)^{N_f}, \qquad \sum_{n=1}^f n N_n = N.
  \ee
  Here, $(n)$ denotes a cyclic permutation of $n$ elements, while the exponent $N_n$ denotes the number of such $n$-cycles present in the conjugacy class.
The centralizer subgroup is of the form
  \be
  C_g = \left(S_{N_1} \right) \times  \left(S_{N_2} \times \ZZ_2 \right)  \ldots \left(S_{N_f} \times \ZZ_f \right).
  \ee
  Here, the factors $S_{N_n}$ permute the $N_n$ cycles $(n)$, while the factor $\ZZ_n$ is generated  by the cyclic permutation within a
  single cycle $n$.

  Each of the twisted sector Hilbert spaces can be written as
  \be 
   \calh_g^{C_g} =\bigotimes_{n>0} S^{N_n} \calh_{(n)}^{\ZZ_n},\label{Hdecomp}
   \ee
   where by $S^{N } \calh$ we mean the
   symmetric tensor product of the Hilbert space $\calh$.
   The Hilbert
 space $\calh_{(n)}$ is obtained by taking $n$ copies of $\calc$ and imposing boundary conditions twisted by the generator $\O$ of the cyclic permutations. More concretely,  
we impose that the  bosonic fields $X$  of the theory obey the boundary conditions
 \be 
 X_i (\s + 2 \p) =  \O X_{i}(\s), \qquad i = 1, \ldots , n .\label{nbc}
 \ee 
 where $\s \sim \s + 2 \p$ is the periodic  coordinate on the cylinder  and $\O$ acts as
 \be 
 \O X_i = X_{i+1}, i = 1, \ldots , n-1 , \qquad \O X_n = X_1.
 \ee
  Note that the  $X_i$ satisfy $ X_i (\s + 2n \p) =
 X_i (\s)$, and can be seen as strands of a `long string'  whose worldsheet is a cylinder with circumference
 $2 \p n $.  The Hilbert space $\calh_{(n)}$ obtained in this way describes the twisted sector of the cyclic orbifold, see \cite{Borisov:1997nc} for more details.

 The space $\calh_{(n)}^{\ZZ_n}$ in (\ref{Hdecomp}) is obtained from $\calh_{(n)}$ by projecting on
 $\O$-invariant states.
In view of (\ref{nbc}), $\O$ acts on  $\calh_{(n)}$ as $e^{2\pi i(L_0- \bar L_0)}$ and 
$\calh_{(n)}^{\ZZ_n}$ therefore consists of those states in which the difference between the left-moving and the right-moving weights is an integer. 
  The partition  functions $Z$ and $\tilde Z$ of the length-$n$ twist Hilbert space  are therefore  \cite{Dijkgraaf:1998zd}
 \bea 
 Z \left( \calh_{(n)}^{\ZZ_n}\right) &=& \sum_{\D , \bar \D } c(\D , \bar \D) \d^{(n)}_{\D-\bar \D} q^{\D\over n} \bar q^{\bar \D \over n}, \\ \tilde  Z \left( \calh_{(n)}^{\ZZ_n}\right) &=& (q \bar q)^{{c \over 24} \left(n-{1\over n}\right)} \sum_{h , \bar h } \tilde c(h , \bar h) \d^{(n)}_{h-\bar h} q^{h\over n} \bar q^{\bar h \over n} .\label{Zntw}
 \eea
In the second identity we used the fact that $\calh_{(n)}^{\ZZ_n}$ carries a Virasoro representation at central charge $n c$. We see from this expression  that  the zero-point energy   in the $n$-cycle twist sector is \cite{Dixon:1986qv} \be {c  \over 24}\left(n - {1\over  n}\right).\label{zeropointbos}\ee 
We  will review the computation  of this zero-point energy  in section \ref{seceomX} below.
 We also  introduced the symbol $\d^{(n)}_{h-\bar h}$ which imposes the  projection on $\ZZ_n$-invariant states: 
 \be 
 \d^{(n)}_{h} = \left\{ \begin{array}{l}  1 {\rm \ if\ } h = 0\ ( {\rm mod\ } n) \\
 0   {\rm \  otherwise} \end{array}\right. .\label{deltan}\ee
From (\ref{Zntw}) we see that the expansion coefficients in the $n$-cycle twist sector, defined as
 \be 
 \tilde Z \left( \calh_{(n)}^{\ZZ_n}\right) = \sum_{h, \bar h} \tilde c_{(n)} (h, \bar h ) q^h \bar q^{\bar h},
  \ee
  are given in terms of the seed coefficients as
  \be
\tilde   c_{(n)} (h, \bar h ) = \tilde c\left( n h - {c \over 24} \left(n^2 -1 \right),  n \bar h - {c \over 24} \left( n^2 - {1}\right)\right) \d^{(1)}_{h - \bar h} \label{coeffsHn}
  	\ee
  	
 The  Hilbert spaces $\calh_{(n)}^{\ZZ_n}$ will play an important role in our proposed description of the bulk dual theory at large $N$, which will contain fields   taking values in  an internal space which is precisely $\calh_{(n)}^{\ZZ_n}$.  
 For the
 examples of interest, where the seed theory is free, we can give an explicit description of $\calh_{(n)}^{\ZZ_n}$ as a Fock space built up with fractional ${1 \over n}$ moded oscillators; this  will be worked out in sections \ref{secscalar} and \ref{sechss}.
 
We can now discuss the partition function $Z_N$  of the symmetric orbifold CFT  $Sym^N \calc$. From the above
 considerations we have
 \be 
 Z_N = \sum_{ \{N_n\} }\prod_n Z \left( S^{N_n}  \calh_{(n)}^{\ZZ_n}\right)  
 \ee
 It is convenient to define a generating function of the $Z_N$, for which one obtains
  \bea
  \calz &\equiv& \sum_N p^N Z_N\\
  &=& \sum_N  \sum_{ \{N_n\} }\prod_n p^{n N_n} Z \left( S^{N_n}  \calh_{(n)}^{\ZZ_n}\right)  \\
  &=& \prod_{n>0} \sum_{N \geq 0} p^{n N} Z \left( S^{N_n}  \calh_{(n)}^{\ZZ_n}\right)\\
  &=& 
 \prod_{n>0} \prod_{\D, \bar \D} \left( 1- p^n  q^{ {\D\over n}} \bar q^{ {\bar \D\over n}}  \right)^{-c(\D, \bar \D) \d^{(n)}_{\D-\bar \D}} .\label{DMVV}
\eea  This is the celebrated DMVV formula \cite{Dijkgraaf:1996xw} (see also \cite{Dijkgraaf:1998zd,Maldacena:1999bp}). 

 \subsection{Large-$N$ limit} 
 We now  want to study the symmetric orbifold CFT at large $N$; since the central charge of the orbifold theory is proportional to $N$ we expect to find a weakly coupled gravity dual in this limit.
  For the large-$N$ limit to exist unambiguously, it is important that the ground state of the seed theory is unique \cite{Belin:2015hwa}. 
  For theories with fermions, we are therefore interested in the  Neveu-Schwarz sector of the symmetric orbifold, while on the  other hand the above steps to derive the DMVV formula generalize straightforwardly to  fermions in the Ramond sector. We defer the treatment of this technical subtlety in fermionic theories to section \ref{sechss}, and focus for now on bosonic theories.   In this subsection, we will take the large-$N$ limit at the level of the  partition function (\ref{DMVV}) and discuss the underlying Hilbert space of states in the next subsection.
  
 Uniqueness of the vacuum means, in terms of the expansion coefficients $c(\D , \bar \D )$, that
  \be 
  c \left( - {c \over 24},  - {c \over 24} \right)=1\label{vacunique}
  \ee
  To extract the coefficient of $p^N$ in  (\ref{DMVV}) at large $N$, we use the trick of \cite{deBoer:1998us}, see also \cite{Keller:2011xi}.  Defining $\tilde p =  p (q \bar q) ^{ - {c\over 24}}$ and using (\ref{vacunique}), 
   we can isolate the contribution from the vacuum and write
  \bea
  \calz &=&
{1 \over 1- \tilde p} 
 \prod_{n>1} \prod_{\D, \bar \D}\hspace{-0.1cm}\,^{'} \left( 1- \tilde p^n    q^{{\D \over n}  + {c n \over 24 }} \bar q^{{\bar \D \over n}  + {cn \over 24 }} \right)^{-c(\D, \bar \D) \d^{(n)}_{\D-\bar \D}}\nonu
  &\equiv& {1 \over 1- \tilde p} R(\tilde p)
  \eea
  where the prime means that we exclude the vacuum contribution with $n=1, \D = \bar \D= -{c \over 24}$. Expanding $R(\tilde p) =\sum_k a_k \tilde p^k$, the coefficient of $\tilde p^N$ is $\sum_{k=0}^N a_k$.
  For large $N$, this coefficient tends to  $\sum_{k=0}^\infty a_k =R(1) $, leading to
  \be
 \tilde  Z_N \approx
  \prod_{n>1} \prod_{\D, \bar \D}\hspace{-0.1cm}\,^{'} \left( 1-    q^{{\D \over n}  +{c n\over 24 }} \bar q^{{\bar \D \over n}  + {cn \over 24 }} \right)^{-c(\D, \bar \D) \d^{(n)}_{\D-\bar \D}}.\label{ZlargeN}
  \ee
  Heuristically, the above manipulations show that the states in the Hilbert space $S^N \calh$ which survive the large-$N$ limit are those  where the majority (order $N$) of the entries in the tensor product is the vacuum,  with sparse (order 1) entries  consisting of excited states.

  \subsection{Single-particle states} \label{sec1part}
 The main observation is now that the large-$N$ result (\ref{ZlargeN})  has the form of a non-interacting multi-particle partition function. 
  Recalling the standard statistical mechanics result that a bosonic single particle excitation 
   with quantum numbers $\D , \bar \D $, upon `multiparticling' gives rise to a factor
  \be 
\left( 1- q^\D \bar q^{\bar \D} \right)^{- 1},
 \ee
 one finds the large-$N$ result (\ref{ZlargeN}) is a multiparticling of a theory with single-particle spectrum captured by the partition function
  \bea
\tilde  Z_{{\rm 1-part.}} = \sum_{n, h, \bar h}\hspace{-0.1cm}\,^{'} (q \bar q)^{{c\over 24}\left( n- {1\over n} \right) }  \tilde c(h, \bar h)  \d^{(n)}_{h-\bar h} q^{{h \over n} } \bar q^{{\bar h \over n} } .\label{Z1part}
  \eea
 where the prime means that we exclude the term with $n=1, h=\bar h =0$.
  
  This  partition function counts the  symmetric orbifold equivalent of `single-trace' excitations in large-$N$ gauge theories.  According to   AdS/CFT duality, the states counted by (\ref{Z1part})  should correspond to the normalizeable modes  of elementary fields in the bulk. The full large-$N$ result (\ref{ZlargeN}) then  arises from including multi-particle bulk states, i.e. second quantization. 
  
  Comparing (\ref{Z1part}) with  (\ref{ZCtilde},\ref{Zntw}) we see that 
  \be \tilde  Z_{{\rm 1-part.}} = \left(\tilde Z (\calh) -1\right) + \sum_{n=2}^\infty  \tilde  Z \left( \calh_{(n)}^{\ZZ_n}\right) \label{Z1part2}
  \ee
 Therefore  the single particle states are in one-to-one correspondence with the states of  $\calh_{(n)}^{\ZZ_n}$, except for the conformal vacuum  which is not to be considered  as a single-particle excitation (hence the minus 1).
 
 Let us spell out this one-to-one correspondence in more detail.
  The first term in (\ref{Z1part}) comes from the  untwisted sector of the orbifold. Every  non-vacuum state $|\psi_{(1)} \rangle \in \calh$  of the seed theory  gives rise to an untwisted sector state 
  \be 
 | \psi_{(1)} \rangle \otimes |0\rangle \times \ldots \otimes |0\rangle + {\rm cyclic\ permutations}\label{untwstates}
 \ee 
  which is  a single-particle state contributing to the first term in (\ref{Z1part2}).
  
  
 For the terms coming from the sum over $n\geq 2$ in (\ref{Z1part2}) the one-to-one correspondence works as follows: each state $|\psi_{(n)} \rangle \in \calh_{(n)}^{\ZZ_n}$ corresponds to a symmetric orbifold state in the  sector twisted by  $(1)^{N-n} (n)$, viz.
  \be
   | \psi_{(n)} \rangle \otimes |0\rangle \times \ldots \otimes |0\rangle + \ldots .\label{1parttwist}
   \ee
  Here, the dots denote that we should also sum over all possible ways  the cycle  $(1)^{N-n} (n)$ can be embedded in $S_N$ \cite{Dixon:1985jw}.
  The states of the form (\ref{1parttwist}) are the ones counted by the $n$-th term in the sum  in (\ref{Z1part2}).

   In summary, we
  showed that the single-particle states are in one-to-one correspondence with the states in the Hilbert space
   \be
   {\bigoplus_{n=1}^\infty}\, ' \calh_{(n)}^{\ZZ_n},
   \ee
   where the prime means that we exclude the vacuum ray in the $n=1$ sector.
   We will often identify the single particle states in the symmetric orbifold with those in $\calh_{(n)}^{\ZZ_n}$, though we should keep in mind that this is a shorthand notation for  states of the form (\ref{untwstates},\ref{1parttwist}).


  \section{The bulk theory}\label{secbulk}
  In this section we discuss the structure of the bulk dual field theory for general  large-$N$ symmetric orbifolds.
    We  begin  by interpreting the result (\ref{Z1part}) for the single-particle spectrum in terms of $AdS_3$ particle multiplets of definite mass and helicity. The linearized theory around $AdS_3$ could be described by a set  of decoupled fields satisfying the appropriate wave equations, but we want to go a step further and unify these fields into a fully gauge invariant description with  the appropriate higher spin symmetry. 
  This symmetry, as we shall  review, can be extracted from the chiral algebra of the seed CFT.
   Imposing higher spin gauge invariance will naturally lead
  us to equations of the form (\ref{eqsintro}), which describe the bulk theory to linear order in the matter fields.
   The higher spin sector in the bulk is  described by a Chern-Simons theory with higher spin gauge symmetry.  The field equations for the matter sector generalize Vasiliev's unfolded description of matter coupled to higher spin gauge fields \cite{Vasiliev:1992gr}. In the pure  $AdS_3$ background,  the auxiliary fields of our unfolded description can be eliminated using the results of \cite{Kessel:2018zqm}, leading back to the description in terms of  decoupled wave equations.
  We also briefly review  the equivalence of our system of equations in the $AdS_3$ background with the standard unfolded description in the literature.
 As in the previous section, we focus here on bosonic symmetric orbifolds, deferring the  subtleties which arise from fermions to section \ref{sechss}. 
  
   \subsection{Linearized theory around $AdS_3$}\label{secspec}
  To begin our discussion of the bulk dual theory, we will analyze how the single particle states counted by (\ref{Z1part}) organize themselves into $AdS_3$ particle multiplets of definite mass and helicity. 
  The global symmetry algebra of $AdS_3$ is  $sl(2, \RR) \oplus 
 \overline{ sl(2, \RR) }$ and particles correspond to unitary representations of this algebra with energy $L_0 + \bar L_0$ bounded below. They are of the type (primary, primary) 
  and we will label them by the  weights of the primary state  as $(h, \bar h )$. The weights $h$ and $\bar h$ must be positive for unitarity. 
   The character of the $(h, \bar h )$ representation is $\chi_h  \bar \chi_{\bar h}$, where the $sl(2, \RR) $ primary characters are given by
    \begin{align}
    \chi_0&= 1, & \nonu
   \chi_h &= {q^h \over 1-q}, &  {\rm for \ }& h>0.
    \end{align}
  Another way to characterize the particle is by its mass $m$ and helicity $\h$, the latter being the charge under  the spatial $U(1)$ rotation group.
  We will also define the `spin' $s$ to be the absolute value of the helicity, $s = |\h|$, the reason being that the standard   field theory description is in terms of a $2 s+1$-dimensional  dimensional tensor under the Lorentz algebra. 
  The relation between the primary weights  $(h, \bar h)$ and these quantum numbers is given by \cite{Aharony:1999ti}
  \be
  m^2 l_{AdS}^2 = (h+ \bar h - s)(h+ \bar h + s-2), \qquad s = |\h| , \qquad \h =  h - \bar h.\label{Adsmass}
  \ee
  We note that the  multiplets of the types $(h,0)$ and $(0,\bar h )$  are short representations, and we will use the terminology that these representations constitute the `massless higher spin sector' in the bulk (even though the spin can be as low as $\half$ for fermions and $1$ for bosons). The long multiplets with both $h$ and $\bar h$ nonzero will be said to constitute the `matter sector'.
  
  To find the particle content in the bulk, we want to write $\tilde Z -1$ and  $Z ( \calh_{(n)}^{\ZZ_n} )$ for $n>1$ as linear combinations of $sl(2, \RR) \oplus 
 \overline{ sl(2, \RR) }$ characters,
  \bea   \tilde Z(\calh) -1 &=& \sum_{h, \bar h} N_{(1)} (h, \bar h) \chi_h  \bar \chi_{\bar h}\\
  Z ( \calh_{(n)}^{\ZZ_n} ) &=& \sum_{h, \bar h} N_{(n)} (h, \bar h) \chi_h  \bar \chi_{\bar h}, \qquad {\rm for\ } n>1,
  \eea
  so that $ N_{(n)} (h, \bar h)$ counts the number of particles with quantum numbers $(h,\bar h)$ coming from the length-$n$ twist sector.
  The fact that such a decomposition is a priori possible, with positive integer coefficients $ N_{(n)} (h, \bar h)$, is guaranteed by the fact that the
  spaces $\calh_{(n)}^{\ZZ_n}$ carry by construction a representation of ${\rm Vir} \oplus \overline{{\rm Vir}}$, and therefore also of its global (or `wedge') subalgebra $sl(2, \RR) \oplus   \overline{ sl(2, \RR) }$.

 To find the coefficients  $ N_{(n)} (h, \bar h)$, it suffices to insert the identity 
  \be 
 q^h =  \chi_{h} -  \chi_{h+1} 
 \label{trick}
 \ee
 into the expression (\ref{Zntw})  for  $\calh_{(n)}^{\ZZ_n}$. Let us discuss the results for the massless higher spin and matter sectors separately.

  The massless higher spin fields in the bulk correspond to chiral CFT excitations with  either $h=0$ or $\bar h=0$; these come exclusively from the untwisted sector, since  the twisted sectors have a nonvanishing left- and right-moving zero-point energy (\ref{zeropointbos}). The partition function which counts these massless states is  the vacuum character of
 the seed theory minus the contribution from the vacuum itself:
  \be 
 \tilde  Z_{\rm chiral}  = \sum_{h>0} \tilde c(h,0) q^h +  \sum_{\bar h>0} \tilde c(0, \bar h) \bar q^{\bar h}
  \label{vacchar}
  \ee
  Using the identity (\ref{trick}) we find the  number of gauge  multiplets of each helicity to be 
  \bea
  N_{(1)} (h, 0) &=& \tilde c(h,0) - \tilde c(h-1,0) \qquad {\rm for \ } h>0 \nonu
  N_{(1)} (0, \bar h) &=& \tilde c(0, \bar h) - \tilde c(0, \bar h-1)  \qquad {\rm for \ }\bar h>0.\label{countchiral}
  \eea   
  where $\tilde c(h, \bar h)$ are expansion coefficients of the seed theory defined in  (\ref{ZCtilde}).
  
 The matter sector in the bulk arises from states with both $h$ and $\bar h$ nonvanishing, which come  both from the untwisted and the twisted sectors. 
  Using once again (\ref{trick}) one arrives at the following   counting formula for the number of multiplets $N_{(n)} (h, \bar h) $ in terms of the coefficients $ \tilde c_{(n)}(h , \bar h)$ given in (\ref{coeffsHn})
  \be
  N_{(n)} (h, \bar h)  = \tilde c_{(n)}(h , \bar h) -\tilde c_{(n)}(h-1 , \bar h) -\tilde  c_{(n)}(h , \bar h-1)+  \tilde c_{(n)}(h-1 , \bar h -1), \ \ \  {\rm for\ }(n, h, \bar h) \neq (1,0,0).\label{countingf}
  \ee

The linearized bulk theory around $AdS_3$ can be described in terms of  decoupled Fronsdal-like\footnote{Note that, though Fronsdal's  equations in three dimensions do not describe any local degrees of freedom, they do capture the relevant boundary excitations \cite{Giombi:2008vd,Gaberdiel:2010ar}.}  (for the massless higher spin sector) and Fierz-Pauli-like 
(for the matter sector) fields. More precisely, since the Fronsdal and Fierz-Pauli equations are parity-invariant, they  describe  representations of the type $(h,0) \oplus (0,h)$ and $(h,\bar h) \oplus (\bar h,h)$ respectively (see \cite{Rahman:2015pzl} for a review). The bulk dual to a parity-invariant theory therefore contains 
	$N_{(1)}(s,0)= N_{(1)}(0,s)$ spin-$s$ Fronsdal fields and $N_{(n)}(h,\bar h)= N_{(n)}(\bar h,h)$ Fierz-Pauli fields  of spin $s$ and mass $m$ given in (\ref{Adsmass}) in the $n$-th twist sector. If the theory is not parity-invariant, the bulk contains fields describing only a single helicity. For the matter sector, the equations for such a field are  a `chiral half' of the Fierz-Pauli system, generalizing the linearized topologically massive spin-2 theory (see \cite{Deser:1981wh,Tyutin:1997yn,Bergshoeff:2009tb} and  eq. (\ref{chiralFP}) below). In the massless higher spin sector, at least for for fermionic fields a  description
	of a single helicity is known in terms of a standard Fronsdal field with a parity-breaking boundary condition, see section 2.6 in \cite{Campoleoni:2017vds}.
	 
  While such a description  of the bulk theory in terms of decoupled wave equations is  adequate for the linearized dynamics around $AdS_3$, it obscures the higher spin gauge invariance which underlies the theory. We therefore want to go a step further and write equations which are manifestly gauge-invariant under the appropriate higher spin algebra. This will have the advantage that we can expand the theory  around any gauge background away from the pure $AdS_3$ background. This  gauge invariant theory captures  much more information about the dual CFT, for example it  should capture  three-point functions with  up to two nonchiral operators.

  \subsection{Higher spin algebra}\label{secHSalg}
As a first step  in constructing a gauge-invariant description, we review in this section how to obtain the higher spin algebra which is expected to govern the bulk theory.  
  Following the insights of \cite{Gaberdiel:2011wb,Gaberdiel:2015mra,Gaberdiel:2015wpo}, this algebra can be derived from the properties of the CFT
as follows. 

 One starts by considering the algebra of chiral operators
  of the symmetric orbifold which, as   in any CFT, organizes  the states of the theory which  fall in representations of this algebra.  At large $N$,  the operators comprising the chiral 
  algebra can be divided into
  `single-trace' and `multi-trace' operators. Focusing on  the left-moving sector,  the single-trace operators are related by the state-operator correspondence to the states in the untwisted sector of the form
    \be 
 \left( \calu(z)|0\rangle \right) \otimes |0\rangle \times \ldots \otimes |0\rangle + {\rm cyclic\ permutations}
  \ee 
  where $\calu (z)$ is a  chiral current of the seed theory (different from the identity). 
  The single-trace chiral currents of the symmetric orbifold are therefore in one-to-one correspondence with the chiral currents of the seed theory:  for any current $\calu (z) $ in  seed theory the corresponding current in the orbifold is
  the symmetric combination
  \be 
  \widehat  \calu (z) = \sum_{i =1}^N \calu_i (z) \label{reprtwist}
  \ee
 where the index $i$ labels the corresponding operator in the $i$-th copy of the seed theory.
  We note that the orbifold chiral algebra, while completely determined by the chiral algebra of the seed theory, is  in general  not isomorphic to it: for example  the  Virasoro central charge in the hatted algebra will be $N$ times larger than that of the seed theory. In   case  the chiral algebra is nonlinear the nonlinear terms will also be different.
   We will denote   by $\cala (\bar \cala)$ the (anti-)chiral algebra
  of the seed theory, and by  $\widehat \cala ( \widehat{\overline{ \cala}})$ the  (anti-)chiral algebra of the symmetric product.
  
     In the known examples of $AdS_3/CFT_2$, the  bulk higher spin gauge algebra is closely related  to the chiral algebra,  $\widehat \cala \oplus \widehat{\overline{ \cala}}$ in the case at hand. Namely, it is the wedge subalgebra $\widehat \cala_0 \oplus \widehat{\overline{\cala_0}}$ spanned by all modes of the chiral currents which annihilate the conformal 
      vacuum state, both when acting from the left and from the right \cite{Gaberdiel:2011wb}. At large values of the central charge\footnote{In the  examples where the seed theory is a free CFT of interest in this paper, the chiral algebra is linear and the there is no need for the additional large $c$ limit.}, these modes can be argued to form a closed Lie subalgebra \cite{Bowcock:1991zk}.  Since $\widehat \cala_0 \oplus \widehat{\overline{\cala_0}}$ is  a Lie algebra rather than a vertex operator algebra, the construction (\ref{reprtwist}) shows that it is isomorphic   to the wedge subalgebra
     $\cala_0 \oplus \overline{\cala_0}$ of the seed theory acting diagonally on  $N$ copies of the seed theory.
     Summarized, the bulk higher spin algebra is isomorphic to  the wedge algebra
     $\cala_0 \oplus \overline{\cala_0}$ of the seed theory, and at large $N$  all states of the symmetric orbifold CFT must come in representations of this algebra.
    
     We can describe the bulk higher spin algebra more concretely as follows. Suppose we have a basis of the  left-chiral algebra $\cala$ of the seed theory consisting of  quasi-primary currents. We label the elements of this basis as $\calu^{(s)}_{\underline{i}} (z)$, where $s$ is the spin and the subscript $\underline{i}$ labels the spin-$s$ operators. Since each chiral spin-$s$ operator corresponds, through the state-operator mapping, to a primary representation of type $(s,0)$ under
     $sl(2,\RR) \oplus \overline{sl(2,\RR)}$,  the number of chiral spin-$s$ operators is   $N_{(1)} {(s,0)}$ determined in  (\ref{countchiral}).
     The modes of the quasiprimaries  satisfy
     \be 
     [ L_m, \left( U^{(s)}_{\underline{i}} \right)_a] = ( m (s-1) - a)\left( U^{(s)}_{\underline{i}}\right)_{m+a}, \qquad {\rm for\ } m = 0, \pm 1 .\label{commsl2}
     \ee 
Since the vacuum is annihilated from the left by all modes $ (U^{(s)}_{\underline{i}} )_a$ with $a > - s$ (see e.g. \cite{Gaberdiel:1999mc} for a derivation),  the generators of the wedge subalgebra $\cala_0$ are  the  modes 
\be
( U^{(s)}_{\underline{i}} )_a \ {\rm for\  } |a|<s .
\ee

Summarized, we have found the following relations between  the CFT and the bulk algebras
$$\begin{array}{lccc}
\blue{
	Sym^N (\calc ):}&\widehat \cala  & \supset & \widehat{\rm Vir} \\& \red{\downarrow} & & \red{\downarrow}\\\blue{\rm bulk\  HS \ algebra:}& \cala_0 & \supset &sl(2, \RR)  . \nonumber
\end{array}$$
Here, the red arrow means restricting to the wedge algebra. 
In the holographic duality \cite{Gaberdiel:2011wb} between 3D Prokushkin-Vasiliev \cite{Prokushkin:1998bq} theory and minimal model CFTs with $\calw_\infty[\l ]$ symmetry, 
 it is known how to invert the red arrows and obtain the 
(large $c$) CFT chiral algebra from the bulk higher spin algebra: this construction is known as (classical) Drinfeld-Sokolov reduction \cite{Drinfeld:1984qv}. 
It would be interesting to examine if this  construction can be generalized to symmetric orbifolds.
   \subsection{Chern-Simons description of the massless sector} \label{secCS}
   Now we move on to the field theory description of the massless higher spins in the bulk. First, we observe that on the CFT side, the chiral operators form a closed subalgebra under the operator product expansion.
   Therefore one expects that the bulk theory allows for a consistent truncation   to the  gauge  sector.
  From the results of the previous subsection, we should be able to describe the massless sector
  by a gauge field taking values in the Lie algebra  $\cala_0 \oplus \overline{\cala_0}$. 
   Furthermore,
   since massless higher spin fields in three dimensions do not carry any local degrees of freedom 
   (though they do describe nontrivial boundary excitations in $AdS_3$), 
  we expect the gauge sector to be described a Chern-Simons theory, as is the case in the known examples of AdS$_3$/CFT$_2$.
   We are therefore led to the following exact field equations describing the theory truncated to the gauge sector:
   \be 
   F = d A + A \wedge A =0, \qquad   \bar F = d \bar A + \bar A \wedge \bar A =0.\label{flatness}
   \ee
   where the one-form gauge fields $A$ and  $\bar A$ take values in the wedge subalgebras $\cala_0$ and   $ \overline{\cala_0}$ respectively. 
   
   Purely gravitational solutions, with all except the spin-2 higher-spin fields turned off, take values in the $sl(2, \RR) \oplus \overline{sl(2, \RR )}$ subalgebra of the higher-spin algebra.
   In particular, the solution which describes the pure  $AdS_3$ solution in global coordinates $\r, x_\pm = \f \pm t$ is given by
   \be
  A_{AdS} = L_0 d \r +\left( e^\r L_1 + {1 \over 4} e^{ - \r} L_{-1} \right) dx_+, \qquad  \bar  A_{AdS} = \bar L_0 d \r - \left(e^\r \bar L_1 +  {1 \over 4} e^{ - \r} \bar L_{-1} \right) dx_- .\label{AdS}
   \ee
   Note that, in our conventions, the $AdS_3$ translation generators $P_m$ and Lorentz generators $M_m$, where  $m = -1,0,1$, are given by 
   \be 
   P_m = L_m + \bar L_{-m}, \qquad M_m =   L_m - \bar L_{-m}.\label{Lor}
   \ee
   The total connection can be written as $A + \bar A = e^m P_m + \o^m M_m$, where 
   \be e^m = \half \left(A^m + \bar A^{-m} \right), \qquad \o^m = \half \left(A^m - \bar A^{-m} \right)
  \ee are the vielbein and spin-connection one forms respectively. We can write $ A_{AdS}$ and $\bar  A_{AdS}$  in pure gauge form
  \be
   A_{ AdS} = g^{-1} dg, \qquad   \bar A_{ AdS} = \bar g^{-1} d \bar g
   \ee 
   with
   \be g = e^{ \half (L_1 + L_{-1}) x_+} e^{ \r L_0}, \qquad \bar g =  e^{- \half (L_1 + L_{-1}) x_-} e^{ \r L_0}.\label{gs}
   \ee
   We should note that  the $AdS$ background possesses a global symmetry which is given by the full higher spin  algebra  $\cala_0 \oplus \overline{\cala_0}$: indeed, it is left invariant by gauge transformations with parameters
   \be 
   \L = g^{-1} \L_0 g, \qquad   \bar \L = \bar g^{-1} \bar \L_0 \bar g,
   \ee
   where $\L_0 $ and $\bar \L_0 $ are arbitrary elements of $\cala_0$ and $\overline{\cala_0}$ respectively.
   
   Using well-known results in the literature one can see that the Chern-Simons system describes the correct massless wave equations in $AdS_3$. Let us restrict attention for the moment to seed theories which are parity invariant,  so that
   the left- and right chiral algebras are isomorphic and 
   for every  bulk representation with quantum numbers $(h,0)$ we have a parity-related one with  quantum numbers $(0,h)$.
   From our analysis (\ref{countchiral}) we expect the bulk theory to describe $N_{(1)}(s,0) = N_{(1)}(0,s)$ massless spin-$s$ Fronsdal fields. Indeed, using the decomposition (\ref{commsl2}) of the adjoint representation of the higher spin algebra in $sl(2,\RR)$ representations we see that (\ref{flatness}) 
   splits in decoupled equations, each of which can be rewritten as a Fronsdal equation using the 
   results of   \cite{Campoleoni:2010zq}. We expect that a similar analysis can be done in the case the seed theory is not parity-invariant.

Based on previous experiences with $AdS_3/CFT_2$ one might hope to go further 
and show, through a careful analysis  of the asymptotic symmetries that, asymptotically,  the higher spin algebra gets `extended beyond the wedge' to  the large-$c$ limit of the full CFT chiral algebra $\widehat \cala \oplus \widehat{\overline{\cala}}$. 
	For this one would need a way to obtain  the chiral algebra  from the wedge subalgebra  analogous  to Drinfeld-Sokolov (DS) reduction, as remarked upon  in the previous subsection, as well as a set of boundary conditions in the Chern-Simons theory which implement DS reduction in the bulk (as was 
	worked out for the $\calw_\infty [\l ]$ algebra in \cite{Henneaux:2010xg,Campoleoni:2010zq,Campoleoni:2011hg,Gaberdiel:2011wb}). 
It would be interesting to make this more explicit.

    \subsection{Unfolded matter field equations}\label{secmattereqs}
    Having discussed the higher spin gauge fields, we now turn to the bulk description of the  matter sector.
    At the linearized level in the matter fields, we  propose a set of bulk field equations which   are  uniquely determined by the following two criteria:
    \begin{enumerate}
    \item They are fully gauge invariant under finite  higher spin gauge transformations.  
    \item Specializing the background to the $AdS_3$ solution (\ref{AdS}), the equations reduce to the appropriate wave equations for  matter particles  with the quantum numbers and multiplicities  derived   in (\ref{countingf}).
    \end{enumerate} 
    
    Let us first present our equations and then comment on how these  criteria are satisfied.  
  In each twisted sector labelled by $n$ we introduce a master field $|C^{(n)} (x)\rangle $ which is a zero-form in the three-dimensional  spacetime and which takes 
  values in an  internal Hilbert space, hence the ket notation. For $n=1$, in the untwisted sector, this Hilbert space is the subspace $\calh_{nonchiral} \subset \calh$ of non-chiral states with both $h,\bar h >0$  (recall that the   purely chiral excitations  are part of the  gauge sector).  For $n>1$,  the internal Hilbert space is taken to be  the
  $n$-cycle twisted Hilbert space $\calh_{(n)}^{\ZZ_n}$.
  
  We propose the field equation describing the massive states in the length-$n$ twist sector to be
  \be
  \left( d + A^{(n)} + \bar A^{(n)}  \right)| C^{(n)}\rangle =0, \qquad  n= 1,2, \ldots \label{masseq}
  \ee
  Here, the superscript $(n)$ in $ A^{(n)}, \bar A^{(n)}$ means that the gauge connections  are  to be taken in the appropriate representation of 
  $\cala_0 \oplus \overline{\cala_0}$ acting on  the $n$-cycle twisted sector.  We will 
  return to this important point  momentarily.

   The equations (\ref{masseq}) simply state that $ |C_{(n)}\rangle$ is a covariantly constant $ \calh_{(n)}^{\ZZ_n}$-valued section associated to the trivial gauge bundle defined by $A, \bar A$. We also note that they are formally  similar to Vasiliev's unfolded equations \cite{Vasiliev:1992gr} describing a massive scalar coupled to the higher spin algebra $hs[\l]$. We will comment on the precise relation between
  our equations and Vasiliev-like unfolded descriptions in the next subsection.  
  
  The equations obviously satisfy the first criterion above with the fields transforming as
  \bea 
 | C^{(n)}\rangle &\to& \L^{(n)} \bar \L^{(n)} | C_{(n)}\rangle \\
   A^{(n)} &\to& \L^{(n)}\left(  A^{(n)}+ d \right)  (\L^{(n)})^{-1}, \qquad  \bar A^{(n)} \to \bar \L^{(n)}\left(  \bar A^{(n)}+ d \right)  ( \bar \L^{(n)})^{-1},
   \eea
  where $\L_{(n)}, \bar \L_{(n)}$ are finite gauge parameters evaluated in the  length-$n$ twist representation.
  We note that the equations describe propagation of matter fields in an arbitrary gauge background, and  should therefore capture holographic matter-matter-current three-point functions \cite{Ammon:2011ua}. We will come back to this point in the Discussion.

 Let us now describe, as promised,  how to evaluate the gauge field $ A^{(n)}$ acting on the $n$-cycle twist sector, so that the  equations  (\ref{masseq}) are well-defined.
  In the  untwisted sector $n=1$, this is  is straightforward, since the generators of $\cala_0 \oplus \overline{\cala_0}$ as defined in section \ref{secHSalg} naturally act on the space  $ \calh_{nonchiral}$ in which $C^{(1)}$ takes values. 
   
    To evaluate $ A^{(n)}$  in the twisted sectors with $n>1$
    we  proceed as follows.
We  recall from (\ref{1parttwist}) that $ \calh_{(n)}^{\ZZ_n}$ is obtained by taking $n$ copies of the seed theory, imposing  boundary conditions twisted by  $\O$ which sends $X_i \to X_{i + 1\ ({\rm mod} \  n)}$ (see (\ref{nbc})) and  projecting on $\O$-invariant states. 
 From 	any operator  $ \calu (z)$ in the chiral algebra $ \cala$ of the seed theory we  construct $	\calu^{(n)} (z) $,  the diagonal generator in $n$-fold copy,
 		\be 
 		\calu^{(n)} (z) = \sum_{i=1}^n  \calu_i (z).
 		\ee
 		Since $\calu^{(n)}$ is invariant under $\O$, its action on $ \calh_{(n)}^{\ZZ_n}$ is well-defined. By the same argument as in section \ref{secHSalg}, the wedge modes $U^{(n)}_a, |a|<s$, where $s$ is the spin of $\calu$, of these operators  generate a Lie algebra isomorphic $\cala_0 \oplus \overline{\cala_0}$. This  then furnishes the sought-after representation in the $n$-cycle twist sector. 
 		In the next sections we will give an explicit algorithm to  work out an explicit representation of the wedge modes $U^{(n)}_a$ in symmetric orbifolds   in terms of the fractional-moded oscillators in the twisted sectors.

   \subsection{Wave equations around $AdS_3$}
Before discussing these examples, we return to  the second property claimed above.  
 We consider the equations (\ref{masseq}) in the $AdS_3$ background (\ref{AdS}) where the connection takes values in the 
  $sl(2, \RR) \oplus \overline{sl(2, \RR )}$ subalgebra of the higher spin algebra. Therefore if we decompose the internal spaces 
  $ \calh_{nonchiral}$ and $ \calh_{(n)}^{\ZZ_n}, n>1$ into irreducible representations of the $sl(2, \RR) \oplus \overline{sl(2, \RR )}$ subalgebra,  the components of $| C_{(n)}\rangle$ in different representations do not mix.
 The equations  (\ref{masseq}) then reduce to a set of  decoupled equations
 \be
  \left( d + A^{(h,\bar h)}_{ AdS} + \bar A^{(h, \bar h)}_{AdS}  \right) | C^{(n,h,\bar h)}_{\underline{i}}\rangle =0, \qquad \underline{i} = 1, \ldots N_{(n)}(h,\bar h) .\label{eqssplit}
 \ee
 where the superscripts in  $A^{(h,\bar h)}$ and  $A^{(h,\bar h)}$ means that we should evaluate the $AdS_3$ connection in the
 $(h, \bar h)$  representation of $sl(2, \RR) \oplus \overline{sl(2, \RR )}$.
 
 Each of these equations is  equivalent to the standard wave equation for a massive particle with quantum numbers $(h, \bar h)$, as was shown in detail  in \cite{Kessel:2018zqm} whose main points we now briefly review. We pick one of the equations in (\ref{eqssplit}) for a particular value of $(h, \bar h)$: 
\be \left( d + A_{ AdS} + \bar A_{ AdS}  \right) | C (x)  \rangle =0.\label{unfolded}\ee Here, we  have dropped the superscripts with the understanding that  $A_{ AdS}, \bar A_{ AdS}$ act on, and 
 $ | C (x)  \rangle $ takes  values in,  the Hilbert space   $\calh_{h, \bar h}$ of the $(h, \bar h )$ 
representation of
 $sl(2, \RR), \overline{sl(2, \RR)}$. Let us first describe the space of solutions to (\ref{unfolded}). The general solution is obtained by picking an arbitrary vector $|C_0 \rangle$ in  $\calh_{h, \bar h}$ as the value of $|C (x) \rangle,$ in the origin of our $AdS_3$ coordinate system and then parallel-transporting it with the $AdS$ connection. One obtains  solutions of the form
\be 
|C (x) \rangle= (g \bar g)^{-1}(x) |C_0 \rangle, \label{scalsols}
\ee 
with the group elements $g, \bar g$ defined in (\ref{gs}) and evaluated here in the $(h, \bar h)$ representation.
 The solution space inherits the Hilbert space structure of $\calh_{h, \bar h}$ and therefore provides us with a field theory realization of $\calh_{h, \bar h}$ in an almost trivial manner. Since the internal space 
 $\calh_{h, \bar h}$ is infinite dimensional, it does so at the cost of introducing an infinite number of auxiliary fields.
 
 These auxiliary fields can be eliminated in terms of a finite set of physical   components which solve a standard matter wave equation.  This can be done in a rather  economical way as follows. In \cite{Kessel:2018zqm} (see also \cite{Iazeolla:2008ix}) we constructed a set of vectors in the representation space $\calh_{h, \bar h}$ which transform in the $2 s +1$-dimensional, with $s = |h- \bar h|$,  `spin-$s$'representation of the $sl(2, \RR )$ subalgebra of Lorentz transformations generated by $M_m$ (see  (\ref{Lor})). In spinor notation, these states can be written as
 \be 
 | \a_1 \ldots \a_{2 s} \rangle, \qquad \a_i = +,-. \label{spins}
 \ee
 These states are understood to be completely symmetric under permutations of the labels  $\a_i$, so that  the number of independent states is $2s+1$.
 Explicit expressions for the states (\ref{spins}) were derived in \cite{Kessel:2018zqm}, which we now adapt in our current conventions. First we relabel the states in the   multiplet as $| s, a \rangle$, with $a = s, s-1,\ldots, -s$. The two labellings are related by  
\be
a = \half \sum_{i=1}^{2 s} \a_i, \qquad  \a_1 \ldots \a_{2s} =   \underbrace{+ \dots +}_{\hbox{\footnotesize{$s+a$}}} \underbrace{- \dots -}_{\hbox{\footnotesize{$s-a$}}}
\ee
The new states satisfy
\be
M_m | s, a \rangle = ( m s + a)| s, a -m \rangle.
\ee
In the case that $\bar h \geq h$, the  states in the Lorentz multiplets are  of the form\footnote{The fields in \cite{Kessel:2018zqm} are related to the  ones here by a reflection  operation, conjugating the second bra vector into a ket. In terms of this reflected field $C$, the field equation takes a form identical to linearized Vasiliev theory, namely $dA + AC - C \bar A =0$.}
\be
| s, a \rangle = \sum_{n\in \NN} v_n (s,a)  | n \rangle_{h} \otimes |n- a- s\rangle_{\bar{h}}, \label{Vsmain}
\ee 
where $ | n \rangle_{h}$ denotes the normalized level $n$ descendant of the primary of weight $h$; more concretely
 \be
| n\rangle_h = \left(n! (2h) (2h +1) \ldots (2h + n-1)\right)^{-\half} (L_{-1})^n |0\rangle_h.\label{Fockbasis}
\ee
The state $ |0\rangle_h $ is the primary state satisfying
\be
L_1 |0\rangle_h =0, \qquad  L_0 |0\rangle_h= h |0\rangle_h, \qquad \,_h\langle 0 |0 \rangle_h = 1.
\ee
Finally, the coefficients $ v_n (s,a)$  in (\ref{Fockbasis}) are given by 
\begin{align}
v_n (s,a)   =&  {(s-a)! \over (2 s)!} \sum_{l=0}^{s+a} (-)^{l} \binom{s+a}{l} \Big( (1-2h-n)_l (-n)_l  (2h-l+n)_{2s} \nonu &  (2-a-s+n)_{s+a-l-1} (1+\D-a+n)_{s+a-l-1} (1-a-s+n) (\D-a+n) \Big)^{1/2} \,, \label{vcoeffs}
\end{align}
where $(x)_n=x (x+1) \dots (x+n-1)$ denotes the Pochhammer symbol.
  Similar expressions can be obtained  in the case that $\bar h < h$ \cite{Kessel:2018zqm}.

 The states (\ref{spins}) are not normalizeable\footnote{This was to be expected, since $\calh_{h, \bar h}$ furnishes a unitary (reducible) representation of the Lorentz algebra, while the states $ | \a_1 \ldots \a_{2 s} \rangle$  transform in a finite-dimensional nonunitary representation.}, but can be shown to have finite overlap with the solutions (\ref{scalsols}).
 We can therefore consider the projection
 \be 
 \f_{\a_1 \ldots \a_{2 s}} (x) = \langle  \a_1 \ldots \a_{2 s} | C (x) \rangle
 \ee
 which is a symmetric multispinor field transforming in the spin-$s$ representation of the Lorentz algebra. It can be shown \cite{Kessel:2018zqm} that this field satisfies the wave equation
 \be 
 \nabla_{\a_1} ^{ \ \b}\f_{\b \a_2 \ldots \a_{2 s}}  + \left({\rm sgn}\, (h-\bar h)\right) (h + \bar h-1) \f_{\a_1 \ldots \a_{2 s}} =0.\label{chiralFP}
 \ee
 This is the standard `topologically massive' wave equation propagating the representation $\calh_{h, \bar h}$. It can be seen as `half' of the more familiar Fierz-Pauli system which  is parity-invariant and propagates $\calh_{h, \bar h} \oplus \calh_{\bar h,  h} $. Therefore, if the seed CFT we started with is parity-invariant, the field equations (\ref{masseq}) can be 
 combined into standard Fierz-Pauli equations for $N_{(n)}(h,\bar h)=  N_{(n)}(\bar h, h)$ fields with mass squared $m^2l_{AdS}^2 =  (h+\bar h) (h + \bar h-2)$ and spin $|h - \bar h|$.   These considerations show that our system of equations satisfies the second criterion of section \ref{secmattereqs}.
 
 The unfolded equation (\ref{unfolded}) can also be related to the  more standard Vasiliev-type unfolded system in $AdS_3$ \cite{Boulanger:2014vya} as follows.
Besides the states (\ref{spins}) in the spin-$s$ representation of the Lorentz algebra, the space $\calh_{h, \bar h}$ also contains states transforming in the spin-$s+1, s+ 2, \ldots $ representations,
 \be 
| \a_1 \ldots \a_{2(s+m)} \rangle, \qquad \a_i = +,-, m \in \NN \label{spint}
\ee
 and these exhaust the finite-dimensional Lorentz representations  contained in $\calh_{h, \bar h}$. The states (\ref{spint}) for all values of $m$ combined fit together in a single irreducible representations of $sl(2, \RR) \oplus \overline{sl(2, \RR)}$.   The corresponding projections
$\f_{\a_1 \ldots \a_{2 (s+ m)}}$ can be combined into a master field
\be 
\F = \sum_{m\in \NN} \f_{\a_1 \ldots \a_{2 (s+ m)}} y^{\a_1} \ldots  y^{\a_{2(s + m)}}
\ee
where the $y^\a$ are auxiliary commuting spinor variables. The field $\F$ so obtained is precisely the master field in the Vasiliev-like unfolded formulation of massive higher spin fields \cite{Boulanger:2014vya}.

\section{Example I: compact  boson}\label{secscalar}
In the next two sections we will apply the above general considerations to two concrete examples.
 The
case we are ultimately interested  in, where the seed theory is the free $\caln = (4,4)$ SCFT on the 4-torus,
will be discussed in section \ref{sechss}, while in this section we start with   the simpler example of a single  compact boson. The emphasis will be on making  the ingredients entering our bulk field equations (\ref{flatness}) and (\ref{masseq}) explicit. We review  the higher spin algebra which has the structure of a `higher spin square' \cite{Gaberdiel:2015mra}, and
give a concrete description of the $n$-cycle twist Hilbert spaces $\calh_{(n)}^{\ZZ_n}$ in terms of oscillators.
We  also present an algorithm  to derive the oscillator expression of the generators of the higher spin algebra acting on these spaces.
Using  this method the massive equations (\ref{masseq}) can be worked out explicitly in any massless higher spin background, at least in principle. To conclude this section we give more details about the bulk spectrum, both in
terms of multiplets of the higher spin algebra and in terms    of $AdS_3$  masses and spins.

\subsection{The seed theory}

We take the seed theory to be that of a single compact boson  $X$  with radius $R$. The Hilbert space $\calh$  is a   Fock space spanned by states of the form
\be 
\a_{- \vec m} \bar \a_{- \vec p} |P, \bar P \rangle.\label{Hstates}
\ee
Here, $|P, \bar P \rangle$ is the ground state  annihilated by the positive oscillator modes and with $\a_0, \bar \a_0$ eigenvalues
\be
P =  {M\over R} + {WR\over 2}, \qquad \bar P =  {M\over R} - {WR\over 2}, \qquad M, W \in \ZZ.
\ee
The integers $M$ and $W$ are the  momentum and winding numbers.
In (\ref{Hstates})  we used a notation
where
 ${\vec m} = (m_1 , m_2, \ldots)$ is a multi-index of positive integers, and
$\a_{-{{\vec m }}}$  stands for $\a_{-m_1} \a_{-m_2} \ldots$.

The partition function  of the seed theory is 
\bea
\tilde Z(\calh ) &=& \left| \prod_{n=1}^\infty (1- q^n)^{-1}\right|^2 \sum_{P, \bar P \in \G_{1,1}}
 q^{\half P^2} \bar q^{\half \bar P^2} \\ 
&=&  \sum_{h, \bar h \in \NN} \sum_{P, \bar P \in  \G_{1,1} } P(h) P(\bar h)  q^{h+ {P^2\over 2}} \bar q^{\bar h +{\bar P^2\over 2}}\label{ZX}
\eea
where  $P(h)$ is the number of partitions of $h$.
In what follows we will mostly assume that  $ R/ \sqrt{2}$ is generic (i.e. irrational) so that the chiral states of the theory  come entirely from the $M=W=0$ sector.  
We will comment on the non-generic case in the next subsection.

\subsection{Higher spin algebra}\label{sechssX}
As we reviewed in section \ref{secHSalg}, the bulk higher spin algebra is isomorphic to the wedge subalgebra of the chiral algebra of the seed theory. For our free scalar example, this higher spin algebra has the interesting structure of a `higher spin square' as we  now briefly review, referring to  \cite{Gaberdiel:2015mra} 
for details.

The  quasi-primary basis elements of the chiral algebra $\cala$ of the free boson at generic ${R \over \sqrt{2}}$ are built from  products of derivatives of $\pa X (z)$.
It is convenient to organize these in an infinite matrix, where each column contains operators with a fixed number of $X$'s, so that as we move down in the rows the  number of derivatives (i.e. the spin)  increases. In the upper left corner, we have the operator $\pa X$ which is the only quasi-primary linear in $X$.  The second column has as its top entry  the stress tensor \be T\equiv  V^2  = -\half : (\pa X)^2:\label{stresst},\ee and contains one quasi-primary operator $V^{2s}$ for each even spin. An explicit basis  is given by \cite{Bakas:1990ry}
\be
V^{2s} = \sum_{k=1}^{2 s -1}{ (-1)^k \over 2 (2 s-1)}  \binom{2 s-1}{k} \binom{2 s-1}{2 s-k} : \pa^k X \pa^{2s-k} X :.
\label{vertgenrs}\ee
These operators generate a subalgebra
$\calw_\infty^{even} [1]$, called in this context the `vertical' $\calw$-algebra. 
The remaining columns form irreducible representations of this algebra. 

Interestingly, the operators in the first row also form a `horizontal' $\calw$-algebra with one quasi-primary operator \be H^s \sim : (\pa X)^s :\ee for each positive integer spin $s$.
These operators generate the algebra $\calw_{1+\infty}[0]$, as can be shown by fermionizing the real boson.

As we argued in section \ref{secHSalg}, by restricting to the wedge modes we obtain the bulk higher spin Lie algebra $\cala_0$ which inherits a similar  structure and is called  `higher spin square';  we will denote it as $hss$. It contains the vertical higher spin subalgebra $hs^{even}[1]$ and the 
horizontal subalgebra $hs[0]$.  It should be noted that the basis of $hss$ so obtained consists of (infinite) linear combinations of normal-ordered monomials
\be 
\a_{- {\vec m}} \a_{\vec p}.\label{monbasis}
\ee
When acting with a $hss$ generator on Fock space states, only a finite number of these monomials contribute and therefore, for most purposes, it should be equivalent to use the monomial basis (\ref{monbasis}) to describe $hss$. 

We end this subsection with a comment on enhanced symmetry points in moduli space. We assumed so far that   ${R \over \sqrt{2}}$ was irrational, in which case the chiral algebra consists of operators with vanishing momentum and winding number. Indeed, recall that turning on momentum and winding contributes
\be 
\D h = \half \left({M\over R} + {W R \over 2}\right)^2, \qquad \D \bar h = \half \left({M\over R} - {W R \over 2}\right)^2, \qquad M,W \in \ZZ
\ee
to the left- and right- moving operator dimensions. We see from this expression that 
at nongeneric points, when $R/ \sqrt{2}$ is a rational number, the chiral algebra gets  enlarged by operators carrying momentum and winding. These can be viewed as `more tensionless' points where  an enlarged
higher spin algebra becomes manifest.  The biggest enhancement occurs at the selfdual point $R = \sqrt{2}$, where the (anti-)chiral algebra gets extended by operators 
carrying $M = \pm  W$ of the form
\be
:e^{i \sqrt{2} M X}(z):, \qquad :e^{-i \sqrt{2} M \bar X}(\bar z):
\ee
This leads to an extra spin-$M^2$ operator of each chirality for each $M \in \ZZ_0$, which for $M = \pm 1$ gives the familiar extension from a $u(1)_1$ to an $su(2)_1$ current algebra.
It would be interesting to explore how the higher spin square structure is extended in the case of  selfdual radius.

\subsection{The Hilbert spaces $\calh_{(n)}^{\ZZ_n}$}\label{secHnX}
We now describe how the $n$-cycle twisted Hilbert spaces  $\calh_{(n)}^{\ZZ_n}$  introduced in section \ref{secdmvv} can be constructed as Fock spaces built up with oscillator modes, see also \cite{David:2002wn}.
We consider $n$ copies of the free boson theory labelled as $X_i, i =1,\ldots , n$,   and introduce the cyclic permutation $\O$ acting as
\be 
\O X_i =  X_{i+1} {\rm \ for\ } i = 1,\ldots , n-1, \qquad \O X_n = X_1.
\ee
The Hilbert space $ \calh_{(n)}$  arises from imposing  $\O$-twisted boundary conditions
 \be 
X_i (\s + 2 \p) = \O X_i (\s) = X_{i + 1 ({\rm mod}\ n)}, \qquad i = 1, \ldots , n .\label{Xibc}
\ee  
The fields $X_i$ can be seen as strands connecting to form a `long string' of circumference $2\p n$.

To deal with the boundary condition (\ref{Xibc}) we make a field redefinition to a  set of fields  $X^{\left({k \over n} \right)}, k = 0, \ldots , n-1$ which diagonalize $\O$:
\be 
X^{\left({k \over n} \right)}   = {1 \over \sqrt{n}} \sum_{j=1}^n e^{-{ 2 \p i k (j-1)\over n}} X_j , \qquad k =0, \ldots , n-1 .\label{Xkn}
\ee
They satisfy
\be 
\O X^{\left({k \over n} \right)}  = e^{2 \p i k \over n} X^{\left({k \over n} \right)}.
\ee
Therefore the mode numbers of $X^{\left({k \over n} \right)}$ have a fractional  part ${k \over n}$; if we map to the sphere  using $z = e^{- i (\s + i \t)}$ the new fields have Laurent expansions 
\be 
\pa X^{\left({k \over n} \right)} = - i \sum_{m \in \ZZ} \a_{m + {k \over n}} z^{- m - {k \over n} - 1}.\label{fractmodes}
\ee
Using  the OPE  of the original fields 
\be 
\pa X_i (z) \pa X_j (0) \sim - {\d_{ij} \over z^2}
\ee
and the identity (recall that $\d^{(n)}_{h}$ was defined in (\ref{deltan}).)
 \be
\d^{(n)}_{h} = {1 \over n} \sum_{k = 0}^{n-1} e^{-{2 \p i h k \over n}}
\ee
we find that the $ X^{\left({k \over n} \right)}$ satisfy the  OPE
\be 
\pa  X^{\left({k \over n} \right)} (z) \pa  X^{\left({l \over n} \right)} (0) \sim - {\d^{(n)}_{k + l} \over z^2}.\qquad {\rm for\ } k,l = 1, \ldots , n-1 .
\ee
By the usual contour argument we find the commutators
 \be 
[\a_{m + {k\over n}}, \a_{p + {l\over n}} ] =\left(m + {k\over n}\right) \d_{k + l,n} \d_{m+p+1,0}.
\ee  
The fractional oscillator modes can also be related to the the modes of the original fields, $X_i$, which are periodic with period $2 \p n$. 
 For example in terms of the modes of $X_1$,
\be
\pa X_1 (z) = - {i \over n} \sum_m \a^1_m z^{-{m\over n} - 1},
\ee
one finds 
\be 
\a_{m + {k \over n}}= {1 \over \sqrt{n}} \a^1_{mn + k}.
\ee
This property plays an important role in the systematic construction of the Hilbert spaces  of cyclic orbifolds  \cite{Borisov:1997nc}.

 Combining all the modes with fractional parts ${k \over n}$ for $k =0,\ldots, n-1$,  we end up with a set of ${1 \over n}$-fractional moded oscillators $\a_{k\over n}, k \in \ZZ,$ with commutation relations
 \be 
 [\a_{k\over n}, \a_{l\over n} ] = {k\over n}  \d_{{k+l \over n},0}.
\ee  
We define ground  states $|P, \bar P \rangle_n$ with momentum  and winding  in the $n$-cycle twist sector which satisfy
\begin{align} 
 \a_{l\over n}   |P, \bar P \rangle_n =& \bar \a_{l\over n}    |P, \bar P \rangle_n =0 , & {\rm for\ } l>&0, \nonu
  \a_0  |P, \bar P \rangle_n =&  {P \over  \sqrt{n}}   |P, \bar P \rangle_n, &
 \bar  \a_0  |P, \bar P \rangle_n =&  {\bar P \over  \sqrt{n}}  |P, \bar P \rangle_n
 \end{align}
 The Hilbert space $\calh_n$ introduced in \ref{secdmvv} is then the Fock space built up by acting with the creation operators on the ground states
 $|P, \bar P \rangle_n$.

We also introduced in section \ref{secdmvv} the Hilbert space  $\calh_n^{\ZZ_n}$ as the  $\O$-invariant subspace of $\calh_n$. This projection keeps only the states for which $L_0- \bar L_0$ is integer, therefore  $\calh_n^{\ZZ_n}$ is spanned by  states
of the form
\be
\a_{-{\vec m \over n}} \bar \a_{-{\vec p \over n }} |P, \bar P \rangle_n, \qquad  {\rm with\ } \sum m_i - \sum p_i + MW = 0\, ({\rm mod \ n})\label{ZnprojX}
\ee
where we again  used a  shorthand notation
where
${\vec m} = (m_1 , m_2, \ldots)$ is a multi-index and
$\a_{-{{\vec m \over n }}}$ stands for $\a_{-{m_1\over n}} \a_{-{m_2\over n}} \ldots$.

\subsection{Twisted  representations of $hss$}\label{seceomX}
In (\ref{masseq}) we proposed a set of equations describing the matter sector in the bulk, which we repeat here for convenience:
 \be
\left( d + A^{(n)} + \bar A^{(n)}  \right)| C^{(n)} (x)\rangle =0, \qquad  n= 1,2, \ldots \label{masseq2}
\ee
Here, $| C^{(n)} (x)\rangle$ is a spacetime zero-form taking values in the Hilbert space $\calh_{(n)}^{\ZZ_n}$, while 
$ A^{(n)} $ and $ \bar A^{(n)}$ are the $hss$ gauge fields evaluated in the appropriate $n$-cycle twist representation acting on $\calh_{(n)}^{\ZZ_n}$. As promised, we now illustrate how to construct the latter
representation explicitly.

Suppose we start  in the seed theory from a wedge mode $U_a, |a|<s$
 of a spin-$s$ chiral current 
\be\calu(z) =  : \calf ( \pa X, \pa^2 X, \ldots) (z) : ,\ee 
i.e.
\be
U_a= {1 \over 2 \pi i} \oint z^{s+a-1 } \calu(z).
\ee
We would like to find its representation $U_a^{(n)}$  on $\calh_{(n)}^{\ZZ_n}$ in terms of the fractional-moded oscillators $\a_{ m \over
 n}$ which we constructed above. 
This can be accomplished using an  algorithm which consists of the following steps:
\begin{enumerate}
\item Construct the symmetric combination in  $X^{\otimes n}$:
\be\calu^{(n)} = \sum_{j=1}^n : \calf ( \pa X_j, \pa^2 X_j, \ldots) : \ee
Being invariant under cyclic permutations, this has well-defined action on $\calh_{(n)}^{\ZZ_n}$.
\item Reexpress the operator in terms of the redefined fields $X^{\left({k \over n} \right)}, k =0, \ldots , n-1 $ on which the cyclic group acts diagonally, using the inverse of (\ref{Xkn}):
\be 
X_j   = {1 \over \sqrt{n}} \sum_{k=0}^{n-1} e^{{ 2 \p i k (j-1)\over n}} X^{\left({k \over n} \right)} , \qquad j =1, \ldots , n-1 
 \ee
  This yields an expression of the form
  \be\calu^{(n)} = : \tilde \calf ( \pa X^{\left({k \over n} \right)}, \pa^2 X^{\left({k \over n} \right)}, \ldots) : \ee
 \item The above expression is conformally normal-ordered as denoted by $: \ :$. 
To get a meaningful oscillator expression, we would like to convert 
it to creation-annihilation normal ordering which we will denote by  $ \NO \ \NO$. 
The method for converting between different orderings is explained in detail in \cite{Polchinski:1998rq}, Ch.2.

We start by computing the difference in normal orderings for the  bilinears:
\be :\pa  X^{\left({k \over n} \right)} (z) \pa  X^{\left({l \over n} \right)} (z') : - \NO  \pa  X^{\left({k \over n} \right)} (z) \pa  X^{\left({l \over n} \right)} (z') \NO= f^{(k,l )} (z, z'), \qquad {\rm for\ } |z|>|z'| \label{propNO}\ee 
and find
\be 
f^{(k,l )} (z, z') = {1- \left( {z'\over z}\right)^{k \over n}\left(1+ {k \over n}\left({z\over z'}-1\right) \right)\over (z-z')^2} {\d_{k + l,n}}
\ee
As a consistency check, we note that this formula is symmetric under simultaneous exchange  of $k$ and $l$ and $z$
and $z'$, as it should. For integer modes, when $k=l=0$, the right hand side vanishes  and we recover the familiar property that in this case both orderings are equivalent.

Using the basic relation  (\ref{propNO}) one can convert between different orderings for more general local operators   using Wick's theorem, i.e.
\be
\calu^{(n)} 
=  \left( \exp\half \int d^2 zd^2 z'\sum_{k,l=0}^{n-1} f^{(k,l )} (z, z') {\d \over \d \left(\pa
	X^{\left({k \over n} \right)} (z) \right) }  {\d \over \d \left(\pa
	X^{\left({l \over n} \right)} (z') \right) } \right) \NO  \tilde \calf \NO \label{relNO}
\ee
Working out the right hand side yields the naive creation-annihilation ordered expression plus correction terms of progressively lower order in the fields.
\item Plug the Laurent expansion of the fields $X^{\left({k \over n} \right)}$, see (\ref{fractmodes}), into the right hand side of (\ref{relNO}) and extract the desired wedge mode using
\be
U_a^{(n)} = {1 \over 2 \pi i} \oint z^{s+a-1 } \calu^{(n)}(z)
\ee
\end{enumerate} 
This then gives the desired oscillator expression for the action of $U_a^{(n)}$ on $\calh_{(n)}^{\ZZ_n}$.
Similar considerations apply to the construction anti-chiral wedge modes in the twisted sectors. 

Let us illustrate the above algorithm  in a few examples. To start with, we want  to find the representation of the $sl(2,\RR)$ wedge modes $L_m, m= -1,0,1$ of the stress tensor $T\equiv V^2 \equiv H^2$ in the length-$n$ twist sector. We find
\bea 
T^{(n)} &\mathrel{\stackrel{\makebox[0pt]{\mbox{\normalfont\tiny 1}}}{=}}& - \half \sum_{j = 1}^n : (\pa X_j )^2 :\\
&\mathrel{\stackrel{\makebox[0pt]{\mbox{\normalfont\tiny 2}}}{=}}& - \half  : \left(\pa	X^{\left({0 \over n} \right)}\right)^2 :
 - \half \sum_{k=1}^{n-1}  : \pa	X^{\left({k \over n} \right)} \pa	X^{\left(1- {k \over n} \right)}  :\\
 &\mathrel{\stackrel{\makebox[0pt]{\mbox{\normalfont\tiny 3}}}{=}}& - \half  \NO \left(\pa	X^{\left({0 \over n} \right)}\right)^2 \NO
 - \half \sum_{k=1}^{n-1} \NO \pa	X^{\left({k \over n} \right)} \pa	X^{\left(1- {k \over n} \right)}  \NO + {1 \over 2 z^2}  \sum_{k=1}^{n-1} {k \over n} \left( 1- {k \over n}\right).
\eea
Here, the number above the equality sign indicates that it is the result  of the corresponding step in the  algorithm. From this expression we find the  result for the wedge modes
 \be 
 L_m^{(n)}  \mathrel{\stackrel{\makebox[0pt]{\mbox{\normalfont\tiny 4}}}{=}} \half \sum_{p \in \ZZ} \NO \a_{-{p \over n}} \a_{m+{p \over n}} \NO + {1 \over 24} \left(n - {1\over n} \right) \d_{m,0}.
 \ee
Hence we recover the well-known zero-point energy in the twisted sectors \cite{Dixon:1986qv} which we anticipated in (\ref{zeropointbos}) and which can also be derived by a variety
 of alternative methods.
 
 Next, let's consider the wedge modes of the  spin-4 operator $V^{4}$ in the vertical subalgebra, see (\ref{vertgenrs}):
 \be 
 V^{4} = - : \pa X \pa^3 X + {3 \over 2}  : (\pa^2 X)^2 :
 \ee
We get
{\bea 
 V^{4(n)} 
&\mathrel{\stackrel{\makebox[0pt]{\mbox{\normalfont\tiny 3}}}{=}}&   \NO -\pa^3	X^{\left({0 \over n} \right)} \pa	X^{\left({0 \over n} \right)} + {3\over 2} \left(\pa^2	X^{\left({0 \over n} \right)}\right)^2 \NO\nonu && 
+ \half \sum_{k=1}^{n-1} \NO -\pa^3	X^{\left({k \over n} \right)} \pa	X^{\left(1-{k \over n} \right)}
 -\pa	X^{\left({k \over n} \right)} \pa^3	X^{\left(1- {k \over n} \right)}+ 3 \pa^2	X^{\left({k \over n} \right)} \pa^2	X^{\left(1-{k \over n} \right)} \NO\nonu
 && + {1 \over 8 z^4} \sum_{k=1}^{n-1} {k\over n} \left(1-  {k\over n} \right) \left(2-5 {k\over n} \left(1- {k\over n} \right) \right)
\eea}
leading to
 \be 
V^{4(n)}_a  \mathrel{\stackrel{\makebox[0pt]{\mbox{\normalfont\tiny 4}}}{=}} \half \sum_{p \in \ZZ} 
\left( a^2 + 5 {p \over n} \left(  {p \over n}-a \right) + 1 \right) \NO \a_{{p \over n}} \a_{a-{p \over n}} \NO + { \left(n^2 -1 \right)^2 \over 48 n^3 } \d_{m,0}.
\ee
This result agrees with the computation in \cite{Gaberdiel:2015wpo} using a  different method.

As a last example, let us work out the representation of the wedge modes of horizontal  spin-3 generator 
\be 
H^{3} = : (\pa X)^3 :
\ee
One finds
\bea 
H^{3,(n)} &\mathrel{\stackrel{\makebox[0pt]{\mbox{\normalfont\tiny 2}}}{=}}&{1 \over \sqrt{n}} \sum_{k,l,m=0}^{n-1}
:  \pa	X^{\left({k \over n} \right)}   \pa	X^{\left({l \over n} \right)}   \pa	X^{\left({m \over n} \right)}  : \d^{(n)}_{k+l+m}\\
 &\mathrel{\stackrel{\makebox[0pt]{\mbox{\normalfont\tiny 3}}}{=}}&{1 \over \sqrt{n}} \sum_{k,l,m=0}^{n-1}
\NO  \pa	X^{\left({k \over n} \right)}   \pa	X^{\left({l \over n} \right)}   \pa	X^{\left({m \over n} \right)}  \NO \d^{(n)}_{k+l+m} - {n^2 -1 \over 4 n^{3\over 2} z^2}  \pa	X^{\left({0 \over n} \right)} 
\eea
leading to the expression for the wedge modes
\be 
H^{3,(n)}_a \mathrel{\stackrel{\makebox[0pt]{\mbox{\normalfont\tiny 4}}}{=}} {i \over \sqrt{n}} \sum_{p,q\in\ZZ}
\NO \a_{p \over n} \a_{q\over n} \a_{a -{ p+q \over n}} \NO + {i \over 4} {n^2-1 \over n^{3\over 2}} \a_a, \qquad a = -2, \ldots , 2
\ee
As a check, we showed that an alternative computation of   the coefficient of the second term, from requiring that the commutation relation (\ref{commsl2}) is satisfied, leads to the same result.

\subsection{More on the spectrum}
To end our discussion of the free scalar example, we would like to comment more on the spectrum of the bulk theory, both from the point of view of the 
higher spin algebra $hss \oplus \overline{hss}$ and from the $AdS_3$ algebra $sl(2,\RR) \oplus \overline{sl(2,\RR )}$.

We start with the gauge sector, whose partition function is
\bea
\tilde Z_{chiral} &=& \sum_{h \in \NN_{>0}} P( h) q^h +  \sum_{\bar h \in \NN_{>0}} P( \bar h) \bar q^{\bar h}\\
&=&  \left( \prod_{n=1}^\infty (1- q^n)^{-1} -1 \right) +  \left( \prod_{n=1}^\infty (1- \bar q^n)^{-1} -1 \right)
\eea
The first term counts purely leftmoving excitations of the form
\be 
\a_{- \vec m} | 0 \rangle_n
\ee From the point of view of the $hss$ these states form an irreducible representation, as one can see using the basis (\ref{monbasis}) 
 (and similarly the right-moving excitations form an irreducible representation of $\overline{hss}$). This representation can be seen as the defining representation and   is also a short representation as explained in \cite{Gaberdiel:2015wpo}. It is referred to  as the `minimal' representation and we will denote it by $min$.
 The gauge sector contains then the representations $(min,1) \oplus (1, \overline{min})$ under
  $hss \oplus \overline{hss}$ .

From the point of view of the $AdS_3$ algebra $sl(2,\RR) \oplus \overline{sl(2,\RR )}$,  we find from (\ref{countchiral}) and (\ref{ZX}) that the number of massless spin-$s$ gauge fields is
\bea
N_{(1)} (1,0)&=& N_{(1)} (0,1) = 1, \\
  N_{(1)} (s,0)&=& N_{(1)} (0,s) = P(s) - P(s-1) \qquad {\rm for\ } s \in \NN_{>1} \label{countingm0X}
\eea
We can also see this decomposition into $sl(2,\RR)$ primary representations more explicitly at the level of Fock space states. For example, one sees that one of the $sl(2,\RR)$ primaries at level $s$ is given by $(\a_{-1})^s |0 \rangle$. These states are related by the state-operator mapping to the generators $H^s (z)$ of the horizontal $\calw_{1+ \infty}[0]$  algebra. At level 4, an extra $sl(2,\RR)$ primary appears and is given by
\be 
\left(2 \a_{-1} \a_{-3} - {3 \over 2}\a_{-2}^2   \right) |0\rangle.
\ee
This state is created by the vertical spin-4 generator $V^4$  in (\ref{vertgenrs}).
Generally, the states at level $s$ form an $P(s)$-dimensional space, while the primary condition that they are annihilated 
by $L_{-1}$ imposes $P(s-1)$ constraints, leading to the counting formula (\ref{countingm0X}). Our  equations (\ref{flatness}) for the gauge sector, linearized around the $AdS_3$ background (\ref{AdS}), decompose  into $  N_{(1)} (s,0)$ decoupled   Fronsdal equations  at each spin $s$ \cite{Campoleoni:2010zq}.
For the lowest spins, labelling a massless spin-$s$ representation as $( s)$, the gauge sector describes
\be
(1)+ (2)+(3)+2 ( 4) + 2 (5) + 4 (6) + 4(7) + 7(8) + 8 (9)+ 12 (10) + 14(11)+21(12)+\ldots
\ee
The asymptotic behaviour at  large spin is, from the well-known asymptotics of $P(s)$,
\be
N_{(1)} (s,0) \sim e^{2\p \sqrt{s \over 6}}.
	\ee
This Cardy-like behaviour of the minimal representation of $hss$ should be contrasted with that of  the analogous minimal representation of the horizontal (vertical) Vasiliev higher spin subalgebra which describe only one massless field of each (even) spin.

Now let us turn to the matter sector, which comes from the non-chiral CFT states in the untwisted sector and from the twisted sectors.
Since the  general analysis including the zero modes is cumbersome due 
to their appearance in the orbifold projection (see (\ref{ZnprojX})),  we will for the rest of this section
restrict our attention  to the subsector with
\be 
P = \bar P =0.
\ee
In this subsector, the seed coefficients factorize:
\be 
\tilde c (h, \bar h) = P(h) P(\bar h).
\ee
In the untwisted sector, the matter contribution is counted by
\be 
\tilde Z_{(1)} = \sum_{h, \bar h \in \NN>0} P(h) P(\bar h) q^h \bar q^{\bar h} = \chi_{min} \bar \chi_{min}
\ee
with $\chi_{min}$ the character of $min$; we see that the  $hss \oplus \overline{hss}$ representation content is 
$(min, \overline{min})$ \cite{Gaberdiel:2015wpo}.
 In the twisted sector  of cycle length $n$, the states are counted by
  \be 
  \tilde Z_{(n)} = (q \bar q)^{{1 \over 24} \left(n-{1\over n}\right)} \sum_{h, \bar h \in \NN>0} P(h) P(\bar h) \d_{h- \bar h}^{(n)} q^{h \over n } \bar q^{\bar h \over n} 
  \ee
Working out the constraint imposed by $\d_{h- \bar h}^{(n)} $ we get a sum of $n$  modulus-square terms
 \be 
\tilde Z_{(n)} = \sum_{k=0}^{n-1} |\chi_{n,k} |^2\label{modsquare}
\ee
with
\be
\chi_{n,k} =\sum_{h \in \NN} P(nh +k) q^{h + {k\over n} + {{1 \over 24} \left(n-{1\over n}\right)}}.
\ee 
This suggests that each $\chi_{n,k}$ is the character of an irreducible representation of $hss$, and that the $n$-cycle twist sector at $P=\bar P=0$ consists of $n$ inequivalent representations under $hss \oplus \overline{hss}$. It would be interesting to study these representations in more detail using the oscillator realization described above.

From the point of view of the  $AdS_3$ algebra $sl(2,\RR) \oplus \overline{sl(2,\RR )}$, the number $N_{(n)}(h,\bar h)$ of $(h, \bar h)$ representations coming  from the $n$-cycle twist sector can be read off from  (\ref{countingf}) and (\ref{coeffsHn}). Using (\ref{ZX}) this reduces  in this case to
  \bea
N_{(n)} (h, \bar h)  &=&  \d_{h- \bar h}^{(1)}  \left( P \left( nh - {1 \over 24} (n^2 -1) \right) -  P \left( n (h-1) - {1 \over 24} (n^2 -1) \right)\right)\times \nonu && \left( P \left( n \bar h - {1 \over 24} (n^2 -1) \right) -  P \left( n (\bar h-1) - {1 \over 24} (n^2 -1) \right) \right).\label{countingmatterX}
\eea
As an illustration we list some low-lying multiplets, labelled by their mass and spin as  $(m^2 l^2_{AdS} , s)$, for the first few twist sectors:
{\footnotesize\bea 
n&=&1:  (0,0) + (4,1)+ (8,0)+(8,2) + 2 (12,3) + (16,1) + 2 (16,4 )  + (24,0) + 2 (24,2)+ 2(36,1)+ 4 (48,0)+ \ldots\nonu
n&=&2:   \left( - {63 \over 64},0\right)+ \left( - {15 \over 64},0\right)+ \left(  {1 \over 64},1\right)+3\left(  {17 \over 64},2\right)+\left(  {17 \over 64},0\right)+ 2\left(  {81 \over 64},1\right)+ 4 \left(  {225 \over 64},2\right)+ \ldots \nonu 
n&=&3: \left( - {80 \over 81},0\right)+ 4 \left( - {56 \over 81},0\right)+  \left( - {32 \over 81},0\right)+ 2  \left(  {4 \over 81},1\right)+4  \left(  {40 \over 81},0\right)+ 4  \left(  {64 \over 81},1\right) + 16   \left(  {208 \over 81},0\right)+ \ldots\nonumber
\eea}

\section{Example II: tensionless AdS$_3$ strings}\label{sechss}
We now turn to our main example of interest, where the seed theory is the $\caln = (4,4)$ SCFT on the four-torus $T^4$, whose symmetric orbifold we denote as $Sym^N (T^4)$. As  mentioned in the Introduction, this theory has recently been shown to be dual to  a tensionless limit of the worldsheet string theory on $AdS_3 \times S_3\times T^4$ with  Neveu-Schwarz flux 
\cite{Eberhardt:2018ouy}.  Therefore  our bulk field equations (\ref{eqsintro})  can be viewed as a first step towards formulating   tensionless string field theory on this background.

In order to work out (\ref{eqsintro})  in this setting, we first generalize the methods of sections \ref{secsymmorb} and \ref{secbulk}
to include the presence of the fermions, making use of supersymmetry and the spectral flow isomorphism.
 We then describe in more detail  the ingredients necessary to evaluate the field equations, starting with  an explicit oscillator realization of 
the single-particle Hilbert spaces $\calh_{(n)}^{\ZZ_n}$ in which the matter fields take values. As in the  free scalar example, the bulk higher spin algebra has a `square' structure and goes under the name of Higher Spin Square. We work out how to evaluate the action of its generators in the twisted sectors.  As a check on our construction we show that    our bulk spectrum decomposes into 
 multiplets of the bulk superalgebra $psu(1,1|2)$, for which we  provide explicit counting formulas.

\subsection{Large-$N$ spectrum of  $\caln = 4$ symmetric orbifolds}\label{secNis4}
In this  subsection, we describe how  the bosonic results of section \ref{secsymmorb} 
generalize to the case where the seed theory
$\calc$ has fermions and `small'
$\caln =(4,4) $ superconformal symmetry\footnote{The results of this   subsection 
	 generalize straightforwardly to  the case of  $\caln = 2$  superconformal seed theories, since we only make use of an $\caln = 2$ subalgebra.}. 
In particular, we aim to  find the single-particle spectrum of the symmetric orbifold  $Sym^N (\calc)$ at large $N$. We are interested in those states which can be interpreted as perturbative excitations above the $AdS$ background; therefore we will restrict attention to the states in the Neveu-Schwartz ($NS$) sector of the orbifold CFT which have energy of order $N^0$ above the vacuum;   Ramond ($R$) sector states  have  an energy gap of order $N$ and will not be considered here. In addition,
	in taking the large-$N$ limit in section \ref{secdmvv} it was important that the the ground state of the seed theory was nondegenerate. In the $R$ sector  it's not clear how to take an unambiguous large-$N$ limit due to the  ground state degeneracy,  as was emphasized in \cite{Belin:2015hwa}.

The small $\caln=4$ superconformal algebra is generated by the stress tensor $T(z)$, an $su(2)$ current algebra at level ${c \over 6}$
with currents $J^\pm (z) , J^3 (z) $, and  fermionic weight-3/2 currents $G^\pm (z),  \tilde G_\pm (z)$. 
 In the $NS$ sector of interest, where the supercurrents are half-integer moded, the wedge subalgebra is the  superalgebra $psu(1,1|2)$. On the bulk side this algebra arises as the algebra of
superisometries of the  $AdS_3$ background. The bosonic
subalgebra  is $sl(2, \RR) \oplus su(2)$, where the latter is the R-symmetry. Its generators are the wedge modes of $T(z)$ and  $J^\pm, J^3 $.
It is standard  to define a  refined partition function keeping keeping track of the R-charges under  $J^3_0$  and $\bar J^3_0$. In the $NS$ sector we define
\be
Z_{NS} (q, \bar q, y, \bar y )= \tr_{NS}  q^{L_0} \bar q^{\bar L_0} y^{2 J^3_0}\bar  y^{2\bar J^3_0} \equiv \sum_{\D, \bar \D, j_3, \bar j_3} c (\D,\bar \D, j_3, \bar j_3 ) q^h \bar q^{\bar h} y^{2j_3} \bar y^{2 \bar j_3}.\label{ZNS}
\ee
As before, we will denote by  $\tilde Z_{NS}$  the partition function multiplied by $(q \bar q)^{c \over 24}$ and the corresponding expansion coefficients by $\tilde c (h,\bar h, j, \bar j )$.

We recall for later convenience that the  $\caln = 4$ algebra possesses a one-parameter family of equivalent realizations which are related by spectral flow\footnote{Apart from the spectral flow parameter $\h$ which labels isomorphic algebras, there is second continuous parameter $\r$  labelling inequivalent $\caln=4$ algebras \cite{Schwimmer:1986mf}. In this work we consider only the $\r=0$ algebra.  }  \cite{Schwimmer:1986mf} which acts  as
\begin{align}
T_\h (z) =& T(z)- {\h \over z} J^3(z) + {c \h^2 \over 24 z^2}, & &\\
J_\h^3(z) =& J^3(z)  - {c \h \over 12 z}, & J_\h^\pm(z) =& z^{\mp \h}  J^\pm(z),\\
G_\h^\pm (z) = & z^{\mp {\h \over 2}} G^\pm (z), &  \tilde G_{\pm  \h} (z) =&  z^{\mp {\h \over 2}}  \tilde G_\pm (z)
\label{spectflow}
\end{align}
In particular, spectral flow maps a state with weight $\D$ and R-charge $j_3$ to a state with
\be
\D_\h = \D - \h j_3 +  {c \h^2 \over 24 }, \qquad j_{3\h}= j_3 - {c \h \over 12 }.\label{spectflow2}
\ee
Under a simultaneous left- and right-moving  spectral flow with parameter $\h$, the partition function transforms as
\be 
Z_\h (q, \bar q, y, \bar y) = (q \bar q)^{c \h^2 \over 24}  (y \bar y)^{-{c \h \over 6}} Z \left(q, \bar q, q^{-{\h\over 2}}y, \bar q^{-{\h\over 2}} \bar y\right).\label{spectflowZ}
\ee

To derive the partition function of the symmetric orbifold, we first observe that the derivation of the DMVV formula in section (\ref{secdmvv}) generalizes straightforwardly to the $R$ sector of the theory, where the fermions in the seed theory are integer-moded
just like the bosons. The  $NS$ sector of the orbifold theory is somewhat less straightforward, for example the naive generalization of the bosonic orbifold projection in the $n$-cycle twist sector, 
$L_0 = \bar L_0 (\ {\rm mod}\ 1) $, is already incorrect in the untwisted sector, as it would project out the half-integer moded fermionic states.  
Therefore we will follow  the strategy of \cite{deBoer:1998us} and start from the DMVV formula  for the partition in the $R$ sector  and  obtain the result for the $NS$ sector partition function by applying the spectral  flow isomorphism (\ref{spectflow}).

We start from the partition function of the  seed theory in the $R$ sector with an insertion of $(-1)^F =  (-1)^{ 2 J^3_0 + 2 \bar J^3_0}  $: 
\be 
Z_R= \tr_{R}  (-1)^{ 2 J^3_0 + 2 \bar J^3_0} q^{L_0} \bar q^{\bar L_0} y^{2 J^3_0}\bar  y^{2\bar J^3_0} \equiv \sum_{\D, \bar \D, j_3, \bar j_3} d (\D,\bar \D, j_3, \bar j_3 ) q^\D  \bar q^{\bar \D} y^{2 j_3} \bar y^{2\bar j_3}\label{ZRdef}
\ee
Following the  arguments  of section \ref{secdmvv} one arrives at   the DMVV generating function for the $RR$ sector of the symmetric orbifold \cite{Maldacena:1999bp}:
\be
\calz_{R}= 
\prod_{n>0} \prod_{\D, \bar \D,j_3, \bar j_3} \left( 1- p^n  q^{ {\D\over n}} \bar q^{ {\bar \D\over n}}   y^{2 j_3} \bar y^{2\bar j_3}  \right)^{- d(\D, \bar \D ,  j_3,\bar  j_3) \d^{(n)}_{\D-\bar \D}} .\label{DMVVRR}
\ee
Applying the spectral flow (\ref{spectflow})  with parameter $\h = 1$, we end up in the $NS$ sector, and from (\ref{spectflowZ})  and recalling that the central charge of $Sym^N (\calc )$ is $N c$ we find that the $ NS$ and $R$ generating functions are related as
\be 
\calz_{NS} (p,q,\bar q, y,\bar y)= \calz_{R} \left(p (q \bar q)^{c \over 24 } (y \bar y)^{- {c \over 6}} ,q,\bar q,q^{-\half}  y, \bar q^{-\half} \bar y\right)
\ee
This gives the $NS$ generating function in terms of the $R$ sector seed coefficients $ d(\D,\bar \D, j_3 ,\bar  j_3)$. Using the spectral flow isomorphism of the seed theory we  can reexpress $\calz_{NS} $ in terms of the $NS$ seed degeneracies
$ c(\D,\bar \D,  j_3, \bar  j_3)$ defined in (\ref{ZNS}).

To subsequently extract the large-$N$ behaviour, we follow the same procedure as in section \ref{secdmvv} and isolate the contribution of the vacuum
with $\D = \bar \D = -{c \over 24}, j_3=\bar j_3 =0$.  The resulting large-$N$ partition function is again of multi-particle form, and one finds for the corresponding single-particle  partition function
\bea
\tilde  Z_{1-part}&=& \sum_{n>0}\tilde Z_{(n)}, \\
\tilde  Z_{(n)}&=&  (q \bar q )^{{c \over 12}(n-1)} (y \bar y)^{{c \over 6}(1-n)} \hspace{-0.2cm}\sum_{h, \bar h, j_3, \bar j_3}\hspace{-0.3cm}\,^{'} \tilde c (h, \bar h, j_3, \bar j_3 )  \d^{(n)}_{h + j_3 -\bar h - {\bar  j_3}} q^{{h\over n} -  j_3 \left(1- {1 \over n} \right)} \bar q^{{\bar h\over n} - \bar  j_3 \left(1- {1 \over n} \right)} y^{2  j_3} \bar y^{2 \bar  j_3} \nonu
&\equiv& \sum_{h, \bar h, j_3, \bar j_3} \tilde c_{(n)}(h, \bar h , j_3, \bar j_3 ) q^h  \bar q^{\bar h} y^{2 j_3} \bar y^{2\bar j_3}
\label{Z1partNS}
\eea
Here, the sum runs over the set of weights and R-charges in the $NS$ sector of the seed theory, and the prime means that for $n=1$ we exclude the vacuum contribution with
$h = \bar h = j_3 =\bar j_3 =0$. 
In other words, the expansion coefficients in the $n$-cycle twisted sector are
\bea
\tilde c_{(n)}(h, \bar h , j_3, \bar j_3 ) &=& \tilde c \left(n h + (n-1)j_3 - {c \over 12} (n-1) , 
n \bar h + (n-1)\bar j_3 - {c \over 12} (n-1),  \right. \nonu && j_3+ {c \over 12} (n-1),
\left. \bar j_3 + {c \over 12} (n-1) \right) \d^{(1)}_{ h + j_3 - \bar h - \bar j_3}\label{cnNS}
\eea

The formula (\ref{Z1partNS}) passes the consistency check that it gives the correct $NS$-sector zero-point energy in the NS sector of
the $n$-cycle  twist sector, namely   $h = \bar h= {c \over 6}(1-n)$,  and the $y-$dependence shows that the ground states in this sector form a degenerate multiplet with spin $j = {c \over 6}(n-1) $ under the $su(2)$ R-symmetry \cite{Lunin:2001pw}. 
It is sometimes useful to reexpress the the result (\ref{Z1partNS}) in terms of the $R$ sector seed coefficients, in terms of which it   takes the somewhat simpler form
\be
\tilde  Z_{(n)}=  (q \bar q )^{{c n \over 12}} (y \bar y)^{-{c n \over 6}} \hspace{-0.2cm}\sum_{\D, \bar \D, j_3, \bar j_3}\hspace{-0.4cm}\,^{'} |d (\D, \bar \D, j_3, \bar j_3 )|  \d^{(n)}_{\D-\bar \D} q^{{\D\over n} -  j_3 } \bar q^{{\bar \D\over n} - \bar  j_3 } y^{2  j_3} \bar y^{2 \bar  j_3} 
\ee
The modulus is included because the $R$ sector coefficients were  defined in (\ref{ZRdef}) with an insertion of  $(-1)^F $, and the prime now means exclusion of the term with $n=1, \D=\bar \D =0, j_3 =\bar j_3 = {c \over 12}$. 

As in our general discussion of section \ref{secdmvv}, the partition function $\tilde Z_{(n)}$  counts states  in one-to-one correspondence with states in the   $n$-cycle twisted  $NS$ sector Hilbert space $\calh_{(n)  NS}^{\ZZ_n}$, and it will be our goal  to give  a concrete oscillator description 
of this space for the $T^4$ theory.

\subsection{The $T^4$ seed theory}
With this goal in mind, we specify from now on to the case where the seed theory  is the free $\caln = (4,4)$ SCFT on the four-torus $T^4$, containing 4  real bosons and 4  real fermions.
We will collect these into two pairs of complex bosons and fermions $X^a, \widetilde X^a, \psi^a, \widetilde \psi^a , a =1,2$ with OPEs
\be 
\pa X^a (z) \widetilde X^b (0) \sim -{\d^{ab}\over z^2},\qquad \psi^a (z) \widetilde \psi^b (0) \sim {\d^{ab}\over z}.
\ee
 The theory possesses a small $\caln =(4,4)$  superconformal algebra of symmetries
 \cite{Ademollo:1976wv} at central charge $c=6$  whose currents are, in our conventions, given by
\begin{align}
T =& - : \pa \widetilde X^a \pa  X^a : - \half (:   \psi^a \pa \widetilde \psi^a : +  :\widetilde \psi^a \pa  \psi^a: ), & 
J^3 =&  \half :  \psi^a \widetilde \psi^a:,\\
J^+ =& \psi^1 \psi^2, & J^- =& \widetilde \psi^1 \widetilde \psi^2,\\
G^+ =&  \psi^1 \pa \widetilde  X^1 + \psi^2 \pa \widetilde  X^2, &  \widetilde G_- =& \widetilde \psi^1 \pa  X^1 +  \widetilde \psi^2 \pa  X^2 \\
G^- =&  \widetilde \psi^2 \pa  \widetilde X^1 -  \widetilde \psi^1 \pa  \widetilde X^2, &  \widetilde G_+ =&  \psi^2 \pa   X^1 -  \psi^1 \pa   X^2,\label{Nis4}
\end{align}
and similarly in the antiholomorphic sector. 

The seed partition function in the $NS$ sector is given by
\be
\tilde Z_{NS}  =\left| \prod_{n=1}^\infty { \left(1+ y q^{n- \half}\right)^2 \left(1+ y^{-1} q^{n- \half}\right)^2
	\over (1 - q^n)^4}\right|^2 \Theta_{T^4} ( q, \bar q),
\label{ZNST4}\ee
where $\Theta_{T^4} $ is a lattice sum coming from the zero modes:
\be 
\Theta_{T^4} ( q, \bar q)= \sum_{P, \bar  P \in \G_{4,4}} q^{ \half P^2} \bar q^{ \half \bar P^2}.
\ee
The expression (\ref{ZNST4}) defines the seed expansion coefficients $\tilde c(h, \bar h, j_3, \bar j_3)$, and from these
and (\ref{Z1partNS}) at $c=6$ we obtain the single-particle spectrum of tensionless strings on $AdS_3$.

\subsection{The Hilbert spaces $\calh_{(n)}^{\ZZ_n}$}\label{secHnsusy}
We now proceed to give a concrete description of the states counted by (\ref{Z1partNS}). As in our general discussion in section \ref{sec1part}, these are in one-to-one correspondence with states in the $n$-cycle twisted, $NS$ sector, Hilbert space which we denote as $\calh_{(n)NS}^{\ZZ_n}$. We now   give a Fock space realization of these Hilbert spaces. 
Our procedure will mirror the derivation of  the partition function in section \ref{secNis4}: we will first consider the
$R$ sector twisted sector Hilbert space, denoted as $\calh_{(n)R}^{\ZZ_n}$, and then perform a spectral flow to the $NS$ sector.

We start by considering $n$ copies of the $T^4$ theory and imposing  boundary conditions  twisted by a cyclic permutation on all the fields  :
\begin{align}
\F^a_i (\s + 2 \p) =& \F^a_{i + 1 ({\rm mod\ } n) } (\s ), & \widetilde  \F^a_i (\s + 2 \p) =& \widetilde \F^a_{i + 1 ({\rm mod\ } n) } (\s ),
\end{align}
where $\F$ stands for either $X$ or $\psi$.
 Next, we make a field redefinition to  diagonalize the action of the twist:
 \begin{align}
\F^{\left( {k\over n} \right) a} =& {1 \over \sqrt{n}} \sum_{j=1}^{n} (\o_n)^{k(j-1)}\F^a_j, &
\widetilde \F^{\left( {k\over n} \right) a} =& {1 \over \sqrt{n}} \sum_{j=1}^{n} (\o_n)^{-k(j-1)} \widetilde \F^a_j, & k=0,\ldots , n-1,
 \end{align}
where $\o_n \equiv e^{-{2 \p i \over n}}$. 
The fields $\F^{\left( {k\over n} \right) a}$ and $\widetilde \F^{\left( {k\over n} \right) a}$ have modes with fractional parts ${k \over n}$ and $-{k \over n}$ respectively. 

In terms of these redefined fields, the $\caln=4$ generators 
$T^{(n)}, J^{3(n)}, J^{\pm(n)}, G^{\pm (n)}, \widetilde G_\pm^{(n)}$ 
take the form 
\begin{align}
T^{(n)} =& -: \pa \widetilde X^{\left( {k\over n} \right) a} \pa  X^{\left( {k\over n} \right) a} : - \half \left(:   \psi^{\left( {k\over n} \right) a} \pa \widetilde \psi^{\left( {k\over n} \right) a} : +  :\widetilde \psi^{\left( {k\over n} \right) a} \pa  \psi^{\left( {k\over n} \right) a}: \right), \nonu
J^{3(n)} =&  \half :  \psi^{\left( {k\over n} \right) a} \widetilde \psi^{\left( {k\over n} \right) a}:, \qquad
J^{+(n)} = \psi^{\left( {k\over n} \right) 1} \psi^{\left(1- {k\over n} \right) 2}, \qquad J^{-(n)} = \widetilde \psi^{\left( {k\over n} \right) 1} \widetilde \psi^{\left(1- {k\over n} \right) 2},\nonu
G^{+ (n)} =&  \psi^{\left( {k\over n} \right) 1} \pa \widetilde  X^{\left( {k\over n} \right) 1} + \psi^{\left( {k\over n} \right) 2} \pa \widetilde  X^{\left( {k\over n} \right) 2}, \qquad\ \ \widetilde G_- = \widetilde \psi^{\left( {k\over n} \right) 1} \pa  X^{\left( {k\over n} \right) 1} +  \widetilde \psi^{\left( {k\over n} \right) 2} \pa  X^{\left( {k\over n} \right) 2}  \nonu
G^{- (n)} =&  \widetilde \psi^{\left( {k\over n} \right) 2} \pa  \widetilde X^{\left(1- {k\over n} \right) 1} -  \widetilde \psi^{\left( {k\over n} \right) 1} \pa  \widetilde X^{\left(1- {k\over n} \right) 2},\ \   \widetilde G_+ =  \psi^{\left( {k\over n} \right) 2} \pa   X^{\left( 1-{k\over n} \right) 1} -  \psi^{\left( {k\over n} \right) 1} \pa   X^{\left(1- {k\over n} \right) 2}, \label{Nis4n}
\end{align}
where a sum over $k=0, \ldots n-1$ is implied.
 The supercurrents $ G^{\pm (n)}(z), \widetilde G_\pm^{(n)}(z)$
constructed in this way are periodic around the origin in the complex plane and therefore this construction gives the $R$ sector of the $n$-cycle twisted theory. 

The $R$ sector Hilbert space  is built by acting with the modes of the fields  $\F^{\left( {k\over n} \right) a}$ and $\widetilde \F^{\left( {k\over n} \right) a}$ on a suitable ground state
$|0 \rangle_{n}^R$. Apart from being annihilated by the bosonic oscillators with  mode number $\geq 0$ we choose it to be annihilated by the following fermionic modes
\be
\psi^a_{m + {k \over n}} |0 \rangle_{n}^R = \widetilde \psi^a_{m + 1- {k \over n}} |0 \rangle_{n}^R =0 \qquad 
{\rm for\ } m \geq 0, k =0,\ldots, n-1 .\label{groundR} 
\ee
In particular, the zero-modes $\psi^a_0$  annihilate the ground state while the  $\widetilde \psi^a_0$ create other ground states degenerate  with $|0 \rangle_{n}^R$. Using the techniques of section \ref{sechssX} one shows that $|0 \rangle_{n}^R$ has  quantum numbers
\be
\D = \bar \D = 0, \qquad j_3 = \bar j_3 = \half \label{groundQNR}
\ee
In terms of the $su(2)$ R-symmetry, $|0 \rangle_{n}^R$ and $\widetilde \psi^1_0 \widetilde \psi^2_0 |0 \rangle_{n}^R$ form a doublet while $\widetilde \psi^1_0 |0 \rangle_{n}^R$ and $\widetilde  \psi^2_0 |0 \rangle_{n}^R$ are singlets.
The $\ZZ_n$ orbifold projection discussed in section \ref{secdmvv} immediately generalizes to theories with fermions in the  $R$  sector  and imposes
\be L_0 = \bar L_0 \ ({\rm mod\ } 1).\label{orbprojR}
\ee
From these considerations, the $R$ partition function in the $n$-cycle twist sector is
\be
Z \left(\calh_{(n)R}^{\ZZ_n} \right) = \sum_{\D,\bar \D,j_3, \bar j_3} |d (\D, \bar \D,j_3, \bar j_3) | q^{\D \over n} 
{\bar q}^{\bar \D \over n} y^{2 j_3} \bar y^{2 \bar j_3}.\label{ZnR}
\ee

To obtain the desired  $n$-cycle twisted Hilbert  space  $\calh_{(n)}^{\ZZ_n}$ in the  $NS$  sector, we perform a spectral flow (\ref{spectflow}) at $c = 6 n$ with $\h = 1$.  
As we see from  (\ref{spectflow}), (\ref{Nis4n}), this transformation is induced by the following transformation of the fields
\be
\psi^{\left( {k\over n} \right)a}_{NS} (z)= z^{- \half} \psi^{\left( {k\over n} \right)a} (z), \qquad \widetilde \psi^{\left( {k\over n} \right)a}_{NS} (z) = z^{ \half} \widetilde \psi^{\left( {k\over n} \right)a}(z).
\ee
 The  $NS$ sector fields so obtained have mode expansions (dropping the subscript $\,_{NS}$ from now on)
\begin{align}
\pa X^{\left( {k \over n} \right) a} =& - i \sum_{m\in\ZZ} \a^a_{m+{k\over n}} z^{-m-{k\over n} -1} , & \pa \widetilde X^{\left( {k \over n} \right) a} =& - i \sum_{m\in\ZZ}\widetilde  \a^a_{m-{k\over n}} z^{-m+{k\over n} -1}, \\
\psi^{\left( {k \over n} \right) a} =&  \sum_{m\in\ZZ} \psi^a_{m+{k\over n}+ \half} z^{-m-{k\over n} -1} , &  \widetilde \psi^{\left( {k \over n} \right) a} =&  \sum_{m\in\ZZ}\widetilde  \psi^a_{m-{k\over n}-\half} z^{-m+{k\over n} }.
\label{modeexpn}
\end{align}

Under the spectral flow, the $R$ sector ground state $|0 \rangle_{n}^R$ defined in (\ref{groundR}) flows to an $NS$ sector ground state $|0 \rangle_{n}^{NS}$ annihilated by the modes\footnote{One can check that the   $R$ sector states $\widetilde \psi^1_0 \psi^2_0 |0 \rangle_{n}^R, \widetilde \psi^1_0 |0 \rangle_{n}^R$ and $\widetilde \psi^2_0 \psi^2_0 |0 \rangle_{n}^R$ flow to  excited half-BPS states.}
\be
\psi^a_{m + {k \over n} + \half} |0 \rangle_{n}^{NS} = \widetilde \psi^a_{m - {k \over n} + \half} |0 \rangle_{n}^{NS} =0 \qquad 
{\rm for\ } m \geq 0, k =0,\ldots, n-1 .\label{groundNS} 
\ee
We should note that, for ${k\over n}\leq \half$, we have  the standard situation that the ground state is annihilated by the positive-moded oscillators.  For ${k\over n}> \half$ however, we see that the positive
modes $\psi^a_{ {k \over n} - \half}$ do not annihilate the ground state\footnote{Naively the $\psi^a_{ {k \over n} - \half}$ modes for ${k\over n}> \half$ could be used to create states of lower energy than the ground states, but the orbifold projection (see (\ref{orbprojNS}) below) projects these out.}  while the negative modes 
 $\widetilde \psi^a_{ \half- {k \over n} }$ do. As a consequence, one sees that $J^-_0$ annihilates the ground state, while  $J^+_0$ contains $n-1$ terms which don't annihilate it, namely
 \be 
 J^+_0 = \sum_{k=1}^{n-1} \psi^2_{ {k \over n} - \half} \psi^2_{\half - {k \over n} } + \ldots
 \ee
 Therefore the chosen ground state $|0 \rangle_{n}^{NS}$ is the lowest weight state in a spin-${n-1 \over 2}$ multiplet under the $su(2)$ R-symmetry. The spectral flow isomorphism gives  the  quantum numbers of $|0 \rangle_{n}^{NS}$  as
 \be 
 h = \bar h = {n-1 \over 2}, \qquad j_3 = \bar j_3 = -  {n-1 \over 2}\label{groundQN}
 \ee
 which is consistent with the analysis of \cite{Lunin:2001pw}. We will also verify (\ref{groundQN}) explicitly in the next subsection.
   The ground state $|0 \rangle_{n}^{NS}$ is furthermore annihilated by the current modes $G^\a_{\half},   \widetilde G_{\a \half}$  and is therefore a half-BPS state from the point of view of the $psu(1,1|2)$ global algebra. 
   
In the sector with nonvanishing momenta and windings on $T^4$ we similarly we define ground states $| P, \bar P \rangle_{n}$ which have appropriate eigenvalues under the bosonic zero-modes.  The Hilbert
space $\calh_{(n)NS}$ is the Fock space built up by acting on these ground states with the modes of the fields
 (\ref{modeexpn}). Combining the sectors with different values of $k$, we end up with the  oscillator modes
 \be 
 \a^a_{m\over n}, \widetilde \a^a_{m\over n}, \left\{ \begin{array}{lll} \psi^a_{{m \over n}}, & \widetilde \psi^a_{{m \over n}}&
 {\rm for\ } n \ {\rm even}\\ \psi^a_{{m + \half  \over n}}, & \widetilde \psi^a_{{m + \half \over n}}&
 {\rm for\ } n \ {\rm odd}
 \end{array}\right., \qquad m \in \ZZ, a = 1,2
 \ee 
 and similarly in the right-moving sector.
 In writing the above used that for  even $n$ the shift by $\pm \half$ in the fermion  mode numbers in (\ref{modeexpn})  can be absorbed in a redefinition of $m$. 

Finally, to obtain the $n$-cycle twisted Hilbert space  $\calh_{(n)NS}^{\ZZ_n}$ we  have to perform the orbifold $\ZZ_n$ projection. The $R$ sector condition (\ref{orbprojR}) flows in the $NS$ sector to the  projection on states with
\be
L_0 + J^3_0 = \bar L_0 + \bar J_0^3 \ ({\rm mod\ }1).\label{orbprojNS}
\ee 
As a consistency check, we perform a spectral flow transformation on the $R$ partition function (\ref{ZnR}) and find that the $NS$ partition is precisely equal to $\tilde Z_{(n)}$ found in (\ref{Z1partNS}),
\be
\tilde Z \left( \calh_{(n)NS}^{\ZZ_n} \right)= \tilde Z_{(n)}.
\ee
This concludes our Fock space description of the single-particle states of $Sym^N (T^4)$ at large $N$.

We would like to end this subsection with a comment on the equivalence between our expression (\ref{Z1partNS}) for the single-particle spectrum and the one derived in \cite{Eberhardt:2018ouy} using a different method (see also \cite{Gaberdiel:2018rqv}).
 In that work, the result was displayed in a spectral flow frame which arises naturally from the worldsheet theory and is different from ours. 
For even $n$, one sees from  (5.9) in \cite{Eberhardt:2018ouy} that  the partition function is displayed in the $R$ sector and is equal to our $\tilde Z \left( \calh_{(n)R}^{\ZZ_n} \right)= (q \bar q)^{n \over 4}  Z \left( \calh_{(n)R}^{\ZZ_n} \right) $.
For odd $n$,  (5.9) in \cite{Eberhardt:2018ouy} comes from building on a ground state with $h = {1\over 4} \left( n - {1 \over n} \right)$ which is a singlet under $su(2)$ (see  (2.11) in  \cite{Lunin:2001pw} for a discussion). Our ground state (\ref{groundNS}) flows to this  state under spectral flow with parameter $\h = {1\over n} -1$. Applying this flow to  (\ref{Z1partNS}) we obtain
\be 
\tilde Z_{(n)} \to (q \bar q)^{n \over 4} \sum_{\D, \bar \D, j_3, \bar j_3} c (\D, \bar \D, j_3, \bar j_3) \d_{ \D+ j_3 - \bar \D - \bar j_3}^{(n)} q^{\D \over n} \bar q^{\bar \D \over n}y^{2 j_3} \bar y^{2 \bar j_3}
\ee
which is indeed equivalent to  (5.9) in \cite{Eberhardt:2018ouy} for odd $n$.

\subsection{Tensionless string field equations}
After these preliminaries we are  ready to give a concrete meaning to our proposed bulk equations
(\ref{eqsintro})
which describe the tensionless string on $AdS_3$ to linear order in the matter fields. We repeat them here
for convenience: 
\bea 
  F = d A + A \wedge A &=&0, \qquad   \bar F = d \bar A + \bar A \wedge \bar A =0\\
  \left( d + A^{(n)} + \bar A^{(n)}  \right)| C^{(n)}\rangle &=&0, \qquad  n= 1,2, \ldots\label{SFTeqs}
\eea

The first line describes the higher spin gauge fields in the bulk. The connections $A (\bar A)$ take values in the wedge subalgebra of the (anti-) chiral algebra of the $T^4$ theory. As in the compact boson example, we will  restrict our attention to the case of generic $T^4$ moduli so that the (anti-)chiral fields come exclusively from the $P = \bar P=0$ sector with vanishing momenta and windings.
As discussed in \cite{Gaberdiel:2015mra}, the chiral algebra $\widehat \cala$ has,   as in the free boson example,  a `square' structure with the vertical algebra given by the supersymmetric $\calw$=algebra $\calw_\infty^{\caln =4}[0]$.
That work also contains a discussion of the candidates for the horizontal subalgebra, which is less clear-cut than in the free boson example. 

The bulk higher spin algebra is the wedge subalgebra of $\hat \cala$,
which in this case goes under the name Higher Spin Square ($HSS$), and contains the single single-boson higher spin algebra $hss$ discussed in section \ref{sechssX} as a subalgebra. The vertical subalgebra is the supersymmetric Vasiliev higher spin-algebra   $shs_2[0]$.
Summarized, we have encountered the following  algebras governing the bulk and boundary theories:
$$\begin{array}{lccccc}
\blue{Sym^N (T^4)} & \widehat \cala  & \supset &\widehat{\calw}_\infty^{\caln =4}[0] & \supset & \caln = 4\ SCA\\
& \red{\downarrow} & & \red{\downarrow} & & \red{\downarrow}\\
\blue {\rm bulk:}& HSS  & \supset & shs_2[0] & \supset & psu(1,1|2)\nonumber\end{array}$$
where the red arrow again means the operation of taking the wedge subalgebra.

The second line in (\ref{SFTeqs}) describes the matter sector, where the zero-form master field $| C^{(n)}(x) \rangle$ takes values in the
Hilbert space $\calh_{(n)}^{\ZZ_n}$ constructed above\footnote{The matter equation in the  untwisted sector $n=1$ was already proposed 
	in \cite{Raeymaekers:2016mmm}.}. The remaining ingredient in that equation is the evaluation of $ A^{(n)}$ and
$\bar A^{(n)}$, the $HSS$-valued gauge fields  in the representation acting on the $n$-th twisted sector. To work out this representation, we can again follow the algorithm described in section \ref{seceomX}. It allows us to find the oscillator expression of any $HSS$ wedge mode in the $n$-cycle twisted sector once we know  
 the difference between normal-ordered and creation annihilation ordered two-point functions with respect to the ground state (\ref{groundNS}). 
One finds, for $ |z|>|z'|$,
\bea 
 :\pa  X^{\left({k \over n} \right)a } (z) \pa  \widetilde X^{\left({l \over n} \right)b} (z') : - \NO  \pa  X^{\left({k \over n} \right)a} (z) \pa \widetilde  X^{\left({l \over n} \right)b} (z') \NO 
  &=& \d^{a,b}\d^{k,l} {1- \left( {z'\over z}\right)^{k \over n}\left(1+ {k \over n}\left({z\over z'}-1\right) \right)\over (z-z')^2}\nonu 
   :\psi^{\left({k \over n} \right)a } (z)  \widetilde \psi^{\left({l \over n} \right)b} (z') : - \NO  \psi^{\left({k \over n} \right)a } (z)  \widetilde \psi^{\left({l \over n} \right)b} (z') \NO 
  &=& \d^{a,b}\d^{k,l} { \left( {z'\over z}\right)^{k \over n}-1 \over z-z'}.
  \label{propNOT4}
  \eea
  
As an example, we use these to work out the oscillator representation of the $psu(1,1|2) \subset HSS$ generators in the $n$-th twist sector. For the wedge modes of $T^{(n)}$ and $J^{3(n)}$ we find
\bea
L_m^{(n)} &=&  \sum_{p \in \ZZ} \NO \a_{{p \over n}}^a \widetilde \a_{m-{p \over n}}^a   \NO  + \sum_r \left( {r \over n} - {m \over 2} \right) \NO \psi^a_{r \over n} \widetilde  \psi^a_{m -{r \over n}} \NO + {n-1 \over 2} \d_{m,0},\\
J_0^{3(n)} &=& \half \sum_r \NO \psi^a_{r \over n} \widetilde  \psi^a_{ -{r \over n}} - {n-1 \over 2}.
\eea
Here, $r $ is a variable which takes values in $\ZZ$ for $n$ even and in $\ZZ + \half$ for $n$ odd.
This calculation also confirms the ground state quantum numbers (\ref{groundQN}). The remaining  $su(1,1|2)$ generators are straightforward, since they don't involve any normal ordering constants; for example we have
\be
G^{+(n)}_{\pm \half} = \sum_{p \in \ZZ} \left( \psi^1_{ \pm \half -{p \over n}} \widetilde \a_{p \over n}^1 + \psi^2_{ \pm \half -{p \over n}} \widetilde \a_{p \over n}^2 \right).
\ee
The other examples discussed in section (\ref{seceomX}) carry over to the $HSS$ case as well, since   they describe wedge modes of currents made of one of the four real bosons of the $T^4$ theory.  

\subsection{Spectrum and supermultiplet decomposition}

We end our discussion of the tensionless string field theory on $AdS_3$ by commenting on its spectrum, both from the point of view of the 
higher spin algebra $HSS \oplus \overline{HSS}$ and from the $AdS_3$ superalgebra $su(1,1|2)  \oplus \overline{su(1,1|2) }$. For simplicity, we will restrict our analysis to the subsector with 
\be 
P = \bar P =0,
\ee
i.e. with vanishing momentum and winding quantum numbers on $T^4$. In this subsector, the expansion coefficients $\tilde c(h, \bar h, j_3, \bar j_3)$ factorize:
\be 
 \tilde c(h, \bar h, j_3, \bar j_3)  =  \tilde c(h,  j_3)  \tilde c( \bar h, \bar j_3),\label{factor}
 \ee
 where $ \tilde c(h,  j_3)$ are the coefficients of the vacuum character of the chiral algebra
 \bea
\tilde  Z_{vac}^{NS} (q,y) &=&  \prod_{n=1}^\infty { \left(1+ y q^{n- \half}\right)^2 \left(1+ y^{-1} q^{n- \half}\right)^2
 	\over (1 - q^n)^4}\\
 &\equiv & \sum_{h,j_3} \tilde c(h,  j_3) q^h y^{2 j_3}.\label{Zvac}
 \eea

Let us first discuss the spectrum from the point of view of the $HSS \oplus \overline{HSS}$ algebra.
The massless higher spin sector in the bulk  comes from the purely (anti-)chiral excitations in the untwisted sector; its partition function is
\be
\tilde Z_{chiral} =
\left( \tilde Z_{vac}^{NS} -1 \right) + \left(\bar {\tilde  Z}_{vac}^{NS}  -1 \right)\label{gaugepartNS}
\ee
As in the single boson example one can show  \cite{Gaberdiel:2015wpo}  that the first  term in
(\ref{gaugepartNS}) is the character of an irreducible `minimal' representation of $HSS$ which we
will denote as $min$. The gauge sector contains then the  $HSS \oplus \overline{HSS}$  representations $(min,1) \oplus (1, \overline{min})$.

As for the matter sector, it is straightforward to see that the untwisted sector contribution forms a single irreducible representation 
 $(min,\overline{min})$ of  $HSS \oplus \overline{HSS}$. The contribution from the $n$-th twist sector can, using (\ref{ZnR}), be written as a sum of $n$ mod-square terms:
 \be
 \tilde Z_{(n)} = \sum_{k=0}^{n} | \chi_{n,k}^2 |\label{charHSS}
 \ee
 where
 \be\chi_{n,k} = \sum_{m \in \NN} \sum_{j_3 \in \NN/2} |d\left(n m +k , j_3 \right)| q^{ m + {k \over  n} - j_3 + {n \over 2}} y^{2 j_3 - n}.
 \ee
 Here, the $d(\D , j_3)$ are  expansion coefficients of the seed theory vacuum character in the $R$ sector
  \bea
 \tilde  Z_{vac}^{R} (q,y) &=& \left( y^\half - y^{-\half} \right)^2  \prod_{n=1}^\infty { \left(1- y q^{n}\right)^2 \left(1- y^{-1} q^{n}\right)^2
 	\over (1 - q^n)^4}\\
 &\equiv & \sum_{h,j_3}  d(h,  j_3)  q^h y^{2 j_3}.
 \eea
 The expression (\ref{charHSS}) is once again suggestive of the $ \chi_{n,k}$ being characters of irreducible $HSS$ representations, in which case this sector would  contain  $n$ inequivalent irreducible representations of $HSS \oplus \overline{HSS}$.

We now turn to the analysis of the spectrum of the bulk theory from point of view of the $AdS_3$ superalgebra $psu(1,1|2)  \oplus \overline{psu(1,1|2) }$.  For this purpose we will decompose the single-particle partition function (\ref{Z1part})  into  $psu(1,1|2)  \oplus \overline{psu(1,1|2) }$ characters.  
We start by briefly reviewing some representations of the 
 $psu(1,1|2) $ algebra, referring to \cite{Ferreira:2017pgt}, Appendix C, for  details. We will discuss in turn the  long and short multiplets of this algebra.\\
{\bf Long multiplets}\\
These are built on an $sl(2,\RR)$ primary of weight $h$, which is also a   highest weight state of $su(2)$ with weight $j <h$.   The corresponding characters ch$_{h,j}$ are  are:
\begin{align}
{\rm ch} _{h,j} &= \chi_{h,j} + 2 \chi_{h+ \half,j+\half}+ 2 \chi_{h+\half,j-1/2}+ \chi_{h+1,j+1}+ 4 \chi_{h+1,j} + \chi_{h+1,j-1} &  \nonu
&+ 2 \chi_{h+ {3\over 2},j+\half}+ 2 \chi_{h+{3\over 2},j-\half}+ \chi_{h+2,j}, & {\rm for\ } j\geq 1\\
{\rm ch} _{h,\half} &= \chi_{h,\half } + 2 \chi_{h+ \half,1}+ 2 \chi_{h+\half,0}+ \chi_{h+1,{3\over 2}}+ 4 \chi_{h+1,\half}  &  \nonu
&+ 2 \chi_{h+ {3\over 2},1}+ 2 \chi_{h+{3\over 2},0}+ \chi_{h+2,\half} & \\
{\rm ch} _{h,0} &= \chi_{h,0 } + 2 \chi_{h+ \half,\half}+ \chi_{h+1,1}+ 3 \chi_{h+1,0}  + 2 \chi_{h+ {3\over 2},\half}+ \chi_{h+2,0} &
\end{align}
Here, $\chi_{h,j} $ is the  $sl(2,\RR) \oplus su(2)$ character
\be
 \chi_{h,j}  = {q^h ( y^{2j+2} -y^{-2j} ) \over (1-q) (y^2 -1)}. 
 \ee 
The explicit formula reads, in all three cases above:
\be
{\rm ch} _{h,j}= {q^h (y^{2 j+2} - y^{-2j } ) \left(1+ \sqrt{q}y \right)^2 \left(1+ \sqrt{q}y^{-1} \right)^2 \over (1-q) (y^2-1)}. 
\ee
{\bf Short multiplets}\\
These are built on an $sl(2,\RR)$ primary of weight $h$, which is also a   highest weight state of $su(2)$ with weight $h$.   These saturate a unitarity bound  and we will denote  the corresponding characters as ch$_{h}$. They are given by  
\begin{align}
{\rm ch} _{h} &= \chi_{h,h} + 2 \chi_{h+ \half,h-\half}+  \chi_{h+1,h-1}, & {\rm for\ } h\geq 1\\
{\rm ch} _{\half } &= \chi_{\half,\half} + 2 \chi_{1,0}\\
{\rm ch} _{0} &= 1
\end{align}
The explicit formula reads, in all three above cases:
\be
{\rm ch} _{h}= {q^h \left(y^{2 h+2}(1 + \sqrt{q} y^{-1})^2  - y^{-2h} ( 1 + \sqrt{q}y)^2 \right) \over (1-q) (y^2-1)}.
\ee

The long and short multiplet  characters are related by the following  useful  formula which  presumably captures the  structure of null vectors in the Verma module of a short representation:
\be
{\rm ch} _{h} = \sum_{m\in \NN} (-1)^m (m+1)  {\rm ch} _{h + {m \over 2} ,h + {m \over 2}}.\label{nullrel}
\ee
{\bf Supermultiplet spectrum}\\
Now we want to decompose the single-particle partition function (\ref{Z1part}) in terms of these characters; for this we need the analogue of the simple bosonic formula (\ref{trick}). Observing that $su(2)$ invariance implies that the partition function is symmetric under $y \to y^{-1}$, it suffices to write terms of the form $q^h ( y^{2j} + y^{- 2j} ) $ as a linear combination of characters.
One checks  the following identities
\bea 
q^h ( y^{2j} + y^{- 2j} ) &=& \sum_{m\in \NN} (-1)^m (m+1)\left(  {\rm ch} _{h + {m \over 2} ,j + {m \over 2}}- 
 {\rm ch} _{h + {m \over 2} ,j - {m \over 2}-1}\right) \qquad {\rm for\ j < h}\nonu
 q^h ( y^{2h} + y^{- 2h} ) &=& {\rm ch} _{h} - \sum_{m\in \NN} (-1)^m (m+1) 
 {\rm ch} _{h + {m \over 2} ,h - {m \over 2}-1} .
\eea
We note that  the second identity follows from the first one and (\ref{nullrel}). Substituting these into the single-particle partition function (\ref{Z1partNS}) one obtains counting functions for the number of supermultiplets in the spectrum. We will denote by $N_{(n)}\left( (h,j), (\bar h ,\bar j) \right) $ the number of multiplets with $psu(1,1|2)  \oplus \overline{psu(1,1|2) }$ quantum numbers $\left( (h,j), (\bar h, \bar j) \right)$ contained in the $n$-cycle twist sector.  We find
{\small\bea
N_{(n)}\left((h,h),(\bar h, \bar h)\right) &=& \tilde c_{(n)} (h,\bar h,h , \bar h)\nonu
N_{(n)}\left((h,h),(\bar h ,\bar j)\right) &=& \sum_{m \in \NN} (-1)^m (m+1) \left( \tilde c_{(n)} \left(h,\bar h - {m \over 2},h , \bar j- {m\over 2} \right)-  \tilde c_{(n)} \left(h,\bar h - {m \over 2},h , \bar j+ {m\over 2}+1 \right)\right)\nonu
N_{(n)}\left((h,j),(\bar h ,\bar h)\right) &=& \sum_{m \in \NN} (-1)^m (m+1) \left( \tilde c_{(n)} \left(h - {m \over 2} ,\bar h,  j- {m\over 2}, \bar h \right)-   \tilde c_{(n)} \left(h - {m \over 2} ,\bar h,  j+ {m\over 2}+1, \bar h \right)\right)\nonu
N_{(n)}\left((h,j),(\bar h ,\bar j)\right) &=& \sum_{m,p  \in \NN} (-1)^{m+p} (m+1)(p+1) \left( \tilde c_{(n)} \left(h - {m \over 2} ,\bar h - {p \over 2},  j- {m\over 2}, \bar j- {p \over 2} \right)\right.\nonu && \left.-  \tilde c_{(n)} \left(h - {m \over 2} ,\bar h - {p \over 2},  j+ {m\over 2}+1, \bar j- {p \over 2} \right) - \tilde c_{(n)} \left(h - {m \over 2} ,\bar h - {p \over 2},  j- {m\over 2}, \bar j+ {p \over 2}+1 \right)\right.\nonu && \left. + \tilde c_{(n)} \left(h - {m \over 2} ,\bar h - {p \over 2},  j+ {m\over 2}+ 1, \bar j+ {p \over 2}+1 \right)  \right)\label{Nssusy}
\eea}where the coefficients $ c_{(n)} (h,\bar h,j_3 , \bar j_3)$ were given in (\ref{cnNS}). Note that only a finite number of terms in these sums are actually nonvanishing. 

The above  formulas hold for general $\caln = (4,4)$ symmetric orbifolds. To specify to the $T^4$ theory in the $P= \bar P=0$ sector,
we use (\ref{factor},\ref{Zvac}). For example, in the gauge sector we find the lowest lying massless supermultiplets to be
\be
\tilde Z_{vac}^{NS} - 1 =   2\, {\rm ch} _{\half} + {\rm ch} _{1}+ 3\, {\rm ch} _{1,0}+ 2\, {\rm ch} _{{3\over 2},\half}+ 10\, {\rm ch} _{2,0}+ 8\, {\rm ch} _{{5\over 2},\half}  +  {\rm ch} _{3,1}+ 29\, {\rm ch} _{3,0} + \ldots
\ee
In the matter sector we  finds  the following lowest lying multiplets in the first few twist sectors: 
\bea
\tilde Z_{(1)} &=& 4\, {\rm ch}_{\half} {\rm \overline{ch}}_{\half} + 2\left(   {\rm ch}_{\half} {\rm \overline{ch}}_{1}+
 {\rm ch}_{1} {\rm \overline{ch}}_{\half} \right) +  {\rm ch}_{1} {\rm \overline{ch}}_{1} + 9\,  {\rm ch}_{1,0} {\rm \overline{ch}}_{1,0} + 6 \left(   {\rm ch}_{\half} {\rm \overline{ch}}_{1,0}+
 {\rm ch}_{1,0} {\rm \overline{ch}}_{\half} \right)\nonu && + 3\left(   {\rm ch}_{1} {\rm \overline{ch}}_{1,0}+
 {\rm ch}_{1,0} {\rm \overline{ch}}_{1} \right) + \ldots \nonu
 \tilde Z_{(2)} &=&  {\rm ch}_{\half} {\rm \overline{ch}}_{\half}  + 2\left(   {\rm ch}_{\half} {\rm \overline{ch}}_{1}+
 {\rm ch}_{1} {\rm \overline{ch}}_{\half} \right) + 4\, {\rm ch}_{1} {\rm \overline{ch}}_{1} + 4\, {\rm ch}_{\half,0} {\rm \overline{ch}}_{\half,0} +8\left(   {\rm ch}_{\half,0} {\rm \overline{ch}}_{1, \half}+
 {\rm ch}_{1,\half} {\rm \overline{ch}}_{1,0} \right)\nonu && + 64\, {\rm ch}_{1,0} {\rm \overline{ch}}_{1,0} +
 8\left(   {\rm ch}_{\half} {\rm \overline{ch}}_{1,0}+
 {\rm ch}_{1,0} {\rm \overline{ch}}_{\half} \right)+ 16 \left(   {\rm ch}_{1,0} {\rm \overline{ch}}_{1}+
 {\rm ch}_{1} {\rm \overline{ch}}_{1,0} \right)+ 16\,  {\rm ch}_{1,\half} {\rm \overline{ch}}_{1,\half} + \ldots \nonu
  \tilde Z_{(3)} &=&  {\rm ch}_{1} {\rm \overline{ch}}_{1}  + {\rm ch}_{{2\over 3},0} {\rm \overline{ch}}_{{2\over 3},0}
  + 4\,  {\rm ch}_{{5\over 6},\half} {\rm \overline{ch}}_{{5\over 6},\half} + 49\,  {\rm ch}_{1,0} {\rm \overline{ch}}_{1,0}+
  7\left(   {\rm ch}_{1} {\rm \overline{ch}}_{1,0}+
  {\rm ch}_{1,0} {\rm \overline{ch}}_{1} \right) +  \ldots \nonu
   \tilde Z_{(4)} &=&  4\,  {\rm ch}_{1,0} {\rm \overline{ch}}_{1,0} + {\rm ch}_{1,\half} {\rm \overline{ch}}_{1,\half}+100\,
   {\rm ch}_{{5 \over 4},0} {\rm \overline{ch}}_{{5 \over 4},0} +20 
   \left(   {\rm ch}_{{5 \over 4},0} {\rm \overline{ch}}_{{5 \over 4},1}+ {\rm ch}_{{5 \over 4},1} {\rm \overline{ch}}_{{5 \over 4},0} \right) \nonu&& + 64\,  {\rm ch}_{{5 \over 4},\half} {\rm \overline{ch}}_{{5 \over 4},\half}
    + 4\,  {\rm ch}_{{5 \over 4},1} {\rm \overline{ch}}_{{5 \over 4},1}+ \ldots
\eea
The five short multiplets  of the type
$ ({\rm ch} _{\half, \half }, \overline{{\rm ch} _{\half, \half }})$, four coming  from the untwisted sector and one from from the 2-cycle twisted sector, play an important role: each contains 4 massless scalars, which together correspond to the 20 moduli of the background \cite{David:2002wn}. 

As a check on our counting formulas (\ref{Nssusy}) we verified that  the above expansions indeed  agree with (\ref{Z1partNS}) to the required order. The fact that the single-particle spectrum neatly decomposes into supermultiplets is a consistency check on our
derivation of the spectrum in section \ref{secNis4}.

\section{Outlook}
In this work, we proposed a  set of field equations for the tensionless string in $AdS_3$ which generalize the linearized Vasiliev equations for conventional higher spin algebras \cite{Vasiliev:1992gr}. We paid particular attention to the description of the twisted sectors and the interplay of our formulation around $AdS$ with wave equations and Vasiliev-like unfolded formulations.
We conclude by listing some open problems and future directions.

\begin{itemize}
\item It would be interesting to analyze further the representations of the $HSS$ furnished by the twisted sector Fock spaces $\calh_{(n)}^{\ZZ_n}$. Also, while the algorithm of section \ref{secHnX} allows us to derive twisted representations of the $HSS$ generators on a case-by-case basis,  it would be good to have a more explicit expression for the twisted generators.
\item Since our field equations capture cubic couplings like $CCA$, it would be a nontrivial consistency check that they give rise to holographic three-point functions matching the boundary CFT. For linearized Vasiliev theory, this was verified in \cite{Ammon:2011ua}, and it would be interesting to rephrase this calculation in the unfolded description used in this work and generalize it to the full Higher Spin Square theory.
\item Though our setup works in principle for general large-$N$ symmetric orbifolds, it would be of interest to explore in more detail those symmetric orbifolds which have known string duals, such as those describing tensionless strings on $AdS_3 \times S^3 \times K_3$ and $AdS_3 \times S^3 \times S^3 \times S^1$ \cite{Giribet:2018ada,Gaberdiel:2018rqv} and, more recently, the tensionful string on $AdS_3 \times S^3 \times T^4$ at $k>1$ \cite{Eberhardt:2019qcl}. 
\item In this paper we only started to explore the interactions in the bulk theory following from making the theory higher-spin gauge invariant; of course this is not expected to be the full story and it would be very interesting to get a handle on further interactions. Here, one should distinguish between to types: firstly, there could be interactions coming  from `formal' deformations of the higher spin algebra \cite{Sharapov:2019vyd}. In the  case of higher spin theories with $hs[\l ]$ gauge symmetry these were argued to be absent; it is not known whether this result persists for the higher spin square. Secondly, interactions can also be generated by field redefinitions. While  the tensionless string field theory isn't expected to be local, it is generally a challenging problem to find the set of  `minimally nonlocal' field redefinition frames in which the correlators of the dual CFT are reproduced.
Another line of attack would be to directly write a  fully interacting theory generalizing the conventional Vasiliev theories, see \cite{Vasiliev:2018zer} for work in this direction.
\end{itemize}
\begin{appendix}

\acknowledgments

I am greatly indebted to Pan Kessel for initial collaboration on this project and for useful feedback throughout. Furthermore I would like to thank 
A. Campoleoni,    E. Skvortsov and M. Vasiliev for   useful comments and discussions, and I am grateful to L. Eberhardt and R. Gopakumar for useful correspondence and comments on the manuscript. This
 research was supported by the Grant Agency of the
Czech Republic under the grant 17-22899S, and by the European Structural and Investment Fund and the Czech Ministry of Education, Youth and Sports (Project CoGraDS - CZ.02.1.01/0.0/0.0/15\_003/0000437). I would also like to thank the
Erwin Schr\"odinger Institute in  Vienna, where part of this work was completed, for hospitality.\\ \ \\
This work is dedicated  to the memory of \v{L}ubom\'ir Li\v{s}tiak.

\end{appendix}

\end{document}